\documentclass[sigconf]{acmart}

\usepackage{times}
\usepackage{helvet}
\usepackage{courier}
\usepackage{amssymb} 
\usepackage{amsmath}
\usepackage{comment}
\usepackage[ruled,vlined,linesnumbered]{algorithm2e}
\usepackage{booktabs}
\usepackage{graphicx}  
\usepackage{ragged2e}
\usepackage{float}

\setcopyright{rightsretained}

\acmConference[GECCO '17]{the Genetic and Evolutionary Computation Conference 2017}{July 15--19, 2017}{Berlin, Germany}
\acmYear{2017}
\copyrightyear{2017}

\begin{document}

\title{Genetic Algorithm for Epidemic Mitigation by Removing Relationships}

\copyrightyear{2017}
\acmYear{2017}
\setcopyright{acmcopyright}
\acmConference{GECCO '17}{July 15-19, 2017}{Berlin, Germany}\acmPrice{15.00}\acmDOI{http://dx.doi.org/10.1145/3071178.3071218}
\acmISBN{978-1-4503-4920-8/17/07}

\author[F. Concatto]{Fernando Concatto}
\affiliation{%
  \institution{Universidade do Vale do Itaja\'i}
  \city{Itaja\'i}
  \state{SC}
  \country{Brazil}
  \postcode{88302-901}
}
\email{fernandoconcatto@gmail.com}

\author[W. Zunino]{Wellington Zunino}
\affiliation{%
  \institution{Universidade do Vale do Itaja\'i}
  \city{Itaja\'i}
  \state{SC}
  \country{Brazil}
  \postcode{88302-901}
}
\email{w.zunino@yahoo.com.br}

\author[L. A. Giancoli]{Luigi A. Giancoli}
\affiliation{%
  \institution{Universidade do Vale do Itaja\'i}
  \city{Itaja\'i}
  \state{SC}
  \country{Brazil}
  \postcode{88302-901}
}
\email{luigi.giancoli@gmail.com}

\author[R. Santiago]{Rafael Santiago}
\affiliation{%
  \institution{Universidade do Vale do Itaja\'i}
  \city{Itaja\'i}
  \state{SC}
  \country{Brazil}
  \postcode{88302-901}
}
\email{rsantiago@univali.br}

\author[L. C. Lamb]{Lu\'is C. Lamb}
\affiliation{%
  \institution{Federal University of \\Rio Grande do Sul}
  \city{Porto Alegre}
  \state{RS}
  \country{Brazil}
  \postcode{91501-970}
}
\email{lamb@inf.ufrgs.br}

\hyphenation{op-tical net-works semi-conduc-tor in-stances ar-ti-fi-cial-ly cre-at-ed}

\begin{abstract}
Min-SEIS-Cluster is an optimization problem which aims at minimizing the infection spreading in networks. In this problem, nodes can be susceptible to an infection, exposed to an infection, or infectious. One of the main features of this problem is the fact that nodes have different dynamics when interacting with other nodes from the same community. Thus, the problem is characterized by distinct probabilities of infecting nodes from both  the same and from different communities. This paper presents a new genetic algorithm that solves the Min-SEIS-Cluster problem. This genetic algorithm surpassed the current heuristic of this problem significantly, reducing the number of infected nodes during the simulation of the epidemics. The results therefore suggest that our new genetic algorithm is the state-of-the-art heuristic to solve this problem.
\end{abstract}



\maketitle

\section{Introduction}


The infection spreading dynamics have long been the subject of research. One of the first known mathematical models of infection spreading was developed by Daniel Bernoulli in 1766. This first model was based on the concepts of susceptible and immune states, the probability of surviving, death-rate, and force of the infection \cite{Dietz2002}. In 1906, Hamer analyzed the behavior of epidemic diseases such as influenza, dengue and measles, concentrating on their regular recurrence \cite{Hamer1906}. In 1910, Ross developed a system of differential equations to model the transmission of malaria \cite{Ross1910}; since then, his contributions have played an important role in the study of epidemics. An additional model that was highly influential in  epidemiology is the work of Kermack and McKendrick, published in 1927 \cite{Kermack1927}. Their model considers that the individuals of the population can be either susceptible, infected or recovered, and utilizes three interconnected differential equations to specify the behavior of the epidemic.

Infection spreading models go beyond diseases. These models can also be applied to culture, information, or behavior contagion \cite{Pastor-Satorras2014, PastoreYPiontti2014}. For example, a model could be applied to maximize the spread of an idea, like a preference, or to understand the mitigation of a social behavior \cite{PastoreYPiontti2014}.

Epidemic models generally involve modeling the population as complex networks, which are composed of individuals (also called nodes or vertices) and the relationships between them, referred to as connections, links or edges. Usually, the notation $G = (V, E)$ is employed to represent a graph, where $V$ is the set of vertices (individuals) and $E$ is the set of edges (connections) between them. Infections can be transmitted from an infected individual to any susceptible individual that is connected to it \cite{Newman2002spread}.

Mathematical models of infection spreading can be transformed into search problems to maximize or minimize the contagion. In network contexts, the optimization is commonly applied in the number of connections between individuals to be removed \cite{Nandi2016}. One of these models is the Min-SEIS-Cluster, introduced in \cite{Santiago2016}.

Min-SEIS-Cluster tries to minimize the spread of an infection by removing a specified number of connections. In this model, each node from the network can be susceptible (S), exposed (E), or infected (I). The model also defines different infection dynamics in subsets of individuals called clusters. For example, nodes from the same family can affect each other with more intensity. In \cite{Santiago2016}, a heuristic based on successive probabilistic simulations of infection dynamics over the instance is presented.

In this paper, we present a new genetic algorithm for the Min-SEIS-Cluster optimization problem. The results show that this new method found solutions with a smaller amount of infected nodes than the previous heuristic.

This paper is organized as following. Section \ref{sec:background} briefly analyzes works related to the problem of mitigating the spread of infections in an epidemic and defines the Min-SEIS-Cluster problem. Section \ref{sec.ga} introduces our novel genetic algorithm, defining details such as solution encoding and genetic operators. Section \ref{sec:results} discusses the results of our genetic algorithm, comparing them with the current heuristic of the Min-SEIS-Cluster problem. Finally, we present concluding thoughts and suggestions for further research.

\section{Background}
\label{sec:background}


The analysis and control of epidemics is a vast topic of research, involving multiple disciplines such as mathematics, computer science, biology and the social sciences. Many epidemic models have been proposed throughout recent history; an overview of these models along with their properties is presented in \cite{Nowzari2016}. Almost all epidemic models consider that the individuals of a population can be in one of many possible states, the most common being susceptible and infected.

Generally, the mitigation of an epidemic is realized through the modification of the structure of the network that represents the population in question. In \cite{Nowzari2016}, the authors claim that the larger the maximum eigenvalue of the adjacency matrix of a network---the more ``tightly connected'' the network is---, the easier it is for infections to spread. They present two problems: finding the set of $n$ nodes and finding the set of $n$ connections to be removed from a network that best minimizes its maximum eigenvalue, thus improving its resistance to an epidemic. Both problems are NP-hard \cite{Nowzari2016}.

Since these problems are computationally difficult, some studies tried to solve them using relaxations or heuristics. Typical strategies to solve the first problem include the usage of node metrics, such as degree or centrality, to select them for removal \cite{Nowzari2016,Holme2002,Miller2007}. The problem of choosing connections from the network has also been widely explored in the literature \cite{Bishop2011,Saha2015,Mieghem2011,Zanette2008}. The work presented in this study and in the original Min-SEIS-Cluster paper focuses on the second problem.

The removal of elements from a network has an impact on the propagation of infections mostly in two known ways. Removing nodes from the network is correspondent to immunizing individuals, making it impossible for them to become infected. Secondly, removing links between nodes is akin to prohibiting them from interacting; this approach is likely to prevent the contact of a non-infected individual with an infected one \cite{Nandi2016}.

In \cite{Marcelino2009,Marcelino2012}, the authors have researched the behavior of influenza infections along with methods that remove edges and nodes from networks. They compared the results over different flight route networks with 500 nodes (representing airports), and their results demonstrated that the methods which removed edges from the networks were the most efficient. They also described that in practical situations, removing nodes is comparable to shutting down an airport, while removing edges is similar to closing flight routes. The latter is more feasible.

In \cite{Enns2012}, a method to remove links from a network based on a quadratically constrained program is proposed. Their approach involves the minimization of the number of susceptible nodes that are connected to an infected node via any path, and considers a constraint on the amount of links that can be removed. Experiments executed over 15-node Erd\H{o}s--Renyi and small-world graphs with 5 removed links indicated that optimal solutions were found in most cases, and tests using scale-free networks with up to 200 nodes demonstrated that their algorithm was superior to methods which utilize statistics derived from the structure of the network, such as edge centrality.

A comparison of four different heuristics for edge removal to mitigate the spread of an infection is presented in \cite{Nandi2016}. The first heuristic is called MinConnect, and minimizes the number of edges between susceptible and infected nodes. The second heuristic is called MinAtRisk and minimizes the number of nodes that have one or more connections with infected nodes. The third one is called MinPaths. This heuristic minimizes paths that can be used by infected nodes to spread the infection. The last heuristic is named MinWPaths, and it minimizes the weight of edges between susceptible and infected nodes. Through computational experiments using up to 200 nodes, the authors concluded that the most efficient heuristic was MinAtRisk.

A study involving the utilization of evolutionary algorithms to devise a strategy to immunize nodes in a network is presented in \cite{Parousis2014}. The authors used the Susceptible-Infected-Removed (SIR) epidemic model and a basic genetic algorithm to identify how many nodes should be vaccinated in a network, considering a cost for each immunized node. Their experiments, which used 500-node Erd\H{o}s--Renyi and Barabasi--Albert graphs, demonstrated that the amount of vaccinated nodes was kept relatively low over the generations of their algorithm.

In \cite{Santiago2016}, the authors have proposed a computational problem called Min-SEIS-Cluster that attempts to minimize the number of infected nodes over time by removing a specific number of edges. This problem also considers that the nodes can be infected with different dynamics in their social groups, so the nodes inside of the same community might have more probability of infecting each other. A heuristic to select which edges should be removed is also reported in the paper. This heuristic is detailed in the coming Section \ref{sec.heuristic}.

\subsection{Previous Heuristic for Min-SEIS-Cluster}
\label{sec.heuristic}

The method used to develop the original heuristic is based on the Monte Carlo concept, and consists in repeatedly generating random solutions and evaluating them until a sufficiently good solution is found.
\label{explanation.mc}

Algorithm \ref{alg.mc} shows the original heuristic used to solve the Min-SEIS-Cluster problem. The variables $bestValue$ and $bestSolution$ store the best solution found during the search. At the start, the $bestValue$ is zero, and the $bestSolution$ is defined as an empty set. In line 3, a loop starts repeating $attempts$ times. Each attempt is a new random selected set of edges that are removed from the original graph $G$. The set $edgesRemoved$ is composed of the edges removed from $G$, so at each iteration of this loop, a new solution is tested. Inside this loop, simulations are performed $replications$ times over the solution. Before the start of each simulation, the sets $susceptibles$, $exposeds$, $infecteds$ are defined. Then for each time $t \in [T]$, the method $simulateInfection$ is executed to simulate the spreading. At the end of the simulation of an entire solution, the value $worstValue$ is updated if it the largest value of a summation of the number of Infected individuals for all $t \in [T]$  is found.

At the end of all replications assigned to a solution, the $worstValue$ is compared to the best-found $worstValue$ of a solution tested in the past ($bestValue$). So, this is  considered as the best solution found (the one that has presented the worst case better than the others) and presents less infected nodes for all $t \in [T]$.

\begin{algorithm}[h]
  \small
  \DontPrintSemicolon
  \caption{Min-SEIS-Cluster heuristic method}
  \label{alg.mc}
  \SetKwInOut{Input}{Input}
  \SetKwInOut{Output}{Output}
 \Input{$G(V,E)$, $replications$, $attempts$, $T$, $k$, $initialInfected$, $clusters$, $\chi$, $\phi$, $\epsilon$, $\lambda$}
  $bestValue$ $\leftarrow$ $\infty$ \;

  $bestSolution$ $\leftarrow$ $\{\}$ \;
  \ForEach{$attempt \in [attempts]$}{
    $edgesRemoved$ $\leftarrow$ $removeRandomEdges(G, k)$\;
    $E'\leftarrow E\backslash edgesRemoved$\;
    $worstValue \leftarrow -\infty$\;
    \ForEach{$rep \in [replications]$}{
      $infecteds$ $\leftarrow$ $initialInfected$\;
      $susceptibles$ $\leftarrow$ $V \backslash initialInfected$\;
      $exposeds$ $\leftarrow$ $\{\}$\;
      $value \leftarrow 0$ \;
      \ForEach{$t \in [T]$}{

        $simulateInfection(G(V, E'), \chi, \phi, \epsilon, \lambda, clusters,$ $  infecteds, susceptibles, exposeds)$ \;
        $value \leftarrow value + |infecteds|$\;
      }
      \If{$value > worstValue$}{
        $worstValue \leftarrow value$  \;
      }
    }
    \If{$bestValue > worstValue$}{
      $bestValue$ $\leftarrow$ $worstValue$ \;
      $bestSolution$ $\leftarrow$ $edgesRemoved$ \;
    }
  }
  \BlankLine
  \Return{$\{bestSolution, bestValue\}$}\;
\end{algorithm}


The $simulateInfection$ method is explained in Algorithm \ref{alg.si}. This procedure is responsible for updating the states of the nodes in the network, and is divided into two phases. The first phase iterates over all Infected individuals. At each iteration of the loop that starts in line $1$, the infected individual is checked if it remains infected (at line 2). If it does not, the infected individual becomes Susceptible.  If the individual remains Infected, it tries to infect all adjacent Susceptible nodes. If the targeted neighbor is in the same cluster of the Infected individual, it is affected by a probability defined by function $\chi$, otherwise by $\phi$. If the target is Infected, it becomes exposed, and the last period in the new status is stored (given by function $\epsilon$). In the second phase, the exposed individuals are checked if they become Infected. If they become Infected, the period of this new status is calculated by the function $\lambda$.

\begin{algorithm}[h]
  \small
  \DontPrintSemicolon
  \caption{Simulate infection}
  \label{alg.si}
  \SetKwInOut{Input}{Input}
  \SetKwInOut{Output}{Output}
 \Input{$G$, $\chi$, $\phi$, $\epsilon$, $\lambda$, $clusters$, $infecteds$, $susceptibles$, $exposeds$, $t$}
  \ForEach{$infected \in infecteds$}{
    \If{$infected.time \leq t$ }{
      $infecteds \leftarrow infecteds \backslash \{infected\}$\;
      $susceptibles \leftarrow susceptibles \cup \{infected\}$\;
      \textbf{continue}\;
    }
    \ForEach{$target \in susceptibles \cap N(infected)$}{
      $chance \leftarrow 0$\;
      \If{$C_{target} = C_{infected}$}{
        $chance \leftarrow \chi_{C_{target}}(t) $\;
      }\Else{
        $chance \leftarrow \phi(t) $\;
      }
      \If{$random() < chance$}{
        $target.time = t+\epsilon(t)$\;
        $susceptibles \leftarrow susceptibles \backslash \{target\} $\;
        $exposeds \leftarrow exposeds \cup \{target\} $\;
      }
    }
  }
  \ForEach{$exposed \in exposeds$}{
    \If{$exposed.time \leq t$ }{
      $exposed.time \leftarrow t+\lambda(t)$\;
      $exposeds \leftarrow exposeds \backslash \{exposed\}$\;
      $infecteds \leftarrow infecteds \cup \{exposed\}$\;
    }
  }
\end{algorithm}

\section{A Novel Genetic Algorithm}
\label{sec.ga}

Genetic algorithms consist in an optimization method that involves the evolution of a population of chromosomes (solutions) through the search space towards the global optimum, using the biologically-inspired operators of crossover, selection, and mutation \cite{Mitchell1998}. Chromosomes are evaluated by a fitness function, which determines its quality and consequently its probability of being selected for reproduction. Genetic algorithms are particularly efficient in finding solutions for problems with large search spaces \cite{Talbi2009}.

With the intention of improving the rate of convergence of the Min-SEIS-Cluster optimization problem, we developed a genetic algorithm applied to the task of finding a set of connections to be removed from the network which minimizes the number of infection events over a period of time. The search space for this problem is extremely large; its size can be calculated by the binomial coefficient $\binom{n}{k}$, where $n$ is the number of connections of the network, and $k$ determines how many connections should be removed (the $k$ constraint). As such, the random search presented in \cite{Santiago2016} can rarely find good solutions.

This section describes the steps we took to develop a genetic algorithm specialized to the Min-SEIS-Cluster problem, specifying two genetic representations of the solution domain, the fitness function and detailing the operators of crossover and mutation.

\subsection{Solution Encoding}

One of the main aspects of a genetic algorithm is how solutions are encoded as chromosomes. The most common encoding is binary, due to its simplicity and historical usage; in this type of encoding, each chromosome is represented by a fixed-length string of bits, whose values can be either 0 or 1 \cite{Mitchell1998}. Binary encodings are also popular because having a small range of possible values for each gene---while still being able to represent the solution naturally---tends to result in a more performant search \cite{Goldberg1989}. However, alternative types of encoding, such as those using integers or real numbers, have also been shown to perform well in some problems \cite{Mitchell1998}.

We have developed two encodings to represent solutions of the Min-SEIS-Cluster problem. The first encoding, referred to as \textbf{GA-Int}, uses integer-valued genes to specify which connections should be removed from the network, each chromosome having a length equal to $k$. In this encoding, the value of each gene ranges from $1$ to $\left |E \right |$, representing the unique identifier of the connection. The second encoding is named \textbf{GA-Bin} and uses binary strings of length $\left |E \right |$ to represent the solution, where a gene having a value of 1 means that the connection correspondent to that locus should be removed, and 0 indicates that the connection should be left intact. The number of ones in the chromosome must be equal to $k$.

The fitness of a solution is calculated through the simulation of the epidemic in the network, using the algorithms presented in Section \ref{sec.heuristic}. The $removeRandomEdges$ procedure in the fourth line of Algorithm \ref{alg.mc} is replaced with the removal of the edges determined by the solution being currently evaluated. The fitness of the solution is equal to the multiplicative inverse of the replication with the highest amount of infections ($worstValue$).

\subsection{Crossover, Mutation and Chromosome Repair}

Crossover is one of the main mechanisms through which genetic algorithms derive their effectiveness in solving problems. The purpose of the crossover operator is to combine characteristics from two or more solutions to create new ones that tend to be closer to the global optimum \cite{Mitchell1998}. Typical methods to implement crossover include single-point crossover, whereby one position from the chromosome is chosen at random, and the genes of the parents before or after the selected position are switched, forming two new solutions. Another commonly used method is the parametrized uniform crossover, which flips the genes at each position individually, based on a predefined probability \cite{Mitchell1998}.

Mutation is a secondary but important operator that prevents certain values from being permanently lost in some particular positions of the chromosomes \cite{Goldberg1989,Mitchell1998}. Mutation is accomplished by changing the value of one or more randomly-chosen genes to a different value. In the case of binary strings, the chosen bit is simply flipped; in other representations, the value might be incremented, decremented or replaced by another random value.

These operators pose some obstacles in the context of the Min-SEIS-Cluster problem. In the case of GA-Int, multiple genes might have the same value, violating the constraint $k$, since less than the specified number of connections would be removed. For GA-Bin, the number of genes possessing a value of $1$ in the binary string might be less or greater than $k$, also causing a violation.

To resolve these issues, we developed a procedure to repair the chromosomes for both genetic representations. This process ensures that the chromosomes respect the constraint $k$ of the Min-SEIS-Cluster problem. Both algorithms consist in removing or inserting genes randomly, depending on the state of the chromosome in relation to $k$. For the GA-Int representation, the procedure first removes the duplicated numbers and subsequently inserts randomly chosen values between $1$ and $\left |E\right|$ that are not already on the chromosome, until its size is equal to $k$. This corrective strategy is formalized in Algorithm \ref{alg:repair_int}.

\begin{algorithm}[h]
\small
\DontPrintSemicolon
\caption{Chromosome repairment procedure (GA-Int)}
\label{alg:repair_int}
\SetKwInOut{Input}{Input}
\SetKwInOut{Output}{Output}
\Input{$chromosome, k, \left |E\right|$}
\BlankLine
$removeRepeatedGenes(chromosome)$ \;
\BlankLine
\While{$\left| chromosome\right| < k$}{
  $number \leftarrow random(1, \left| E\right|)$ \;
  \If{$number \notin chromosome$}{
    $chromosome \leftarrow chromosome \cup \{number\}$ \;
  }
}
\BlankLine
\Return{$chromosome$} \;
\end{algorithm}

The procedure for the GA-Bin representation works in a similar fashion. The algorithm acts by randomly replacing zeros with ones if the binary string has less ones than $k$, and replaces ones with zeros otherwise. If the chromosome is in accordance with the constraint, the algorithm terminates. This procedure is presented in Algorithm \ref{alg:repair_bin}.

\begin{algorithm}[h]
\small
\DontPrintSemicolon
\caption{Chromosome repairment procedure (GA-Bin)}
\label{alg:repair_bin}
\SetKwInOut{Input}{Input}
\SetKwInOut{Output}{Output}
\Input{$chromosome, k$}
\BlankLine
\While{$countOnes(chromosome) \neq k$}{
  $i \leftarrow random(1, length(chromosome))$ \;
  \If{$countOnes(chromosome) > k$}{
    $chromosome_{i} \leftarrow 0$ \;
  }
  \Else{
    $chromosome_{i} \leftarrow 1$ \;
  }
}
\BlankLine
\Return{$chromosome$} \;
\end{algorithm}

\section{Results and Analysis}
\label{sec:results}

This section presents the results and analysis of the experiments performed by using the classic heuristic proposed in \cite{Santiago2016} and the two versions of our novel genetic algorithm. First, we describe the real-life networks used in our experiments, detailing their source and structure. Next, the parameters for the experiments are specified. Finally, the data collected through the experiments is presented and analyzed.

\subsection{Datasets}
\label{subsec.datasets}

The networks used in our experiments are detailed in Table \ref{tab.instances}, and were obtained in \cite{vlado}. In the following, the datasets and the communities used for each network are detailed. These communities were obtained by the CM+LNM heuristic (Coarsening Merger + Local Node Moving) from \cite{Santiago2017}. The applied exposition and infection times were 2 and 4 respectively.

Strike and Sawmill are networks where the nodes represent the employees of a sawmill. These employees are asked to describe in which frequency they talk to other colleagues on a five-point scale rating. The pair of employees that marked more than three points are connected by an edge in the network. In both networks, there are employees who speak only English or Spanish. Also, they might work in different sections of the facility, resulting in a community structure.




Karate is a network composed of the members of Zachary's karate club. The nodes represent each member of the karate club. Each edge represents the friendship between two members. The club was split due to a conflict between two members of the club, forming two communities.


Korea 1 and 2 are networks where the nodes represent women of two villages in the Republic of Korea. Two nodes are connected by an edge if two women (two nodes) discussed their family planning. In Korea 1, the family-planning program was successful, while in Korea 2 it was not widely adopted. Women who were members of the Mother's Club in their respective villages discussed more family-planning methods.





Dolphins is a network where each node is a dolphin. The association between two dolphins is represented with an edge. The group of dolphins was observed for 7 years.


In the Polbooks instance, each node is a book about politics in the United States, and an edge between two books represents frequent co-purchasing in an online store.

Adjnoun is a network of common adjectives and nouns in David Copperfield, a novel written by Charles Dickens. Football represents the network of American football games between colleges during the Fall of 2000, and Jazz is a collaboration network of jazz musicians, compiled by a research group at the Universidad Rovira i Virgili, in Spain.


\begin{table}[h]
\renewcommand{\arraystretch}{1.0}
\caption{Details of the graphs used for the experiments, specifying the number of nodes and edges. The instances were obtained in \cite{vlado}.}
\centering
\label{tab.instances}
\resizebox{2.3in}{!}{
\begin{tabular}{lrrr}
\toprule
\multicolumn{1}{l}{\textbf{Dataset}} &\multicolumn{1}{l}{\textbf{Nodes}} &\multicolumn{1}{l}{\textbf{Edges}} &\multicolumn{1}{l}{\textbf{Communities}}\\
\cmidrule(lr){1-1}\cmidrule(lr){2-2}\cmidrule(lr){3-3}\cmidrule(lr){4-4}
Strike & 24 & 38 & 4\\
Karate & 34 & 78 & 2\\
Korea1 & 35 & 69 & 10\\
Korea2 & 35 & 84 & 10\\
Sawmill & 36 & 62 & 5\\
Dolphins & 62 & 159 & 7\\
Polbooks & 105 & 441 & 6\\
Adjnoun & 112 & 425 & 9\\
Football & 115 & 613 & 11\\
Jazz & 198 & 2742 & 3\\
\bottomrule
\end{tabular}}
\end{table}

\subsection{Specification of the experiments}

To verify the effectiveness of our proposed genetic algorithm, we performed extensive computational experiments using the original Monte Carlo heuristic in comparison with the novel GA-Int and GA-Bin techniques. For each network, a problem was defined as $10\%$ of its nodes being initially infected, 100 time steps, $replications$ set to 20 and $attempts$ to 300. For the constraint $k$, we experimented with three different values: $0.1\cdot |E|$, $0.3\cdot |E|$ and $0.5\cdot |E|$. All tests were performed considering internal infection with the probability of 15\% and with probability of 5\% between nodes of different clusters.

In the experiments employing our genetic algorithm, we used a parametrized uniform crossover with a probability of exchange at each position equal to $0.5$ and a crossover rate of $0.7$. The mutation operator was slightly different for each genetic representation. In GA-Int, a mutated gene had its value replaced by a random value between 1 and $\left| E\right|$, while in GA-Bin the selected bit was flipped. In both cases, the rate of mutation was set to $0.1$. To select individuals for reproduction, a binary tournament selection operator was used, with a probability of selecting the fitter individual equal to $0.7$.

Both GA-Int and GA-Bin were tested with 300 generations, a value equal to the number of attempts of the original heuristic. To obtain a better insight into their rates of convergence, two different population sizes were specified: 10 and 100.

Since both the original heuristic and the two versions of the genetic algorithm are stochastic search procedures, we collected 10 samples of each experiment to improve the confidence of the data. The results gathered are presented in the next subsection, along with their detailed analysis.

\subsection{Analysis of results}

The analysis of results is divided into two parts. The first part describes the best results obtained by each heuristic. The second part describes the evolution of the heuristic iterations. They are used to compare the methods tested over the datasets reported in Section \ref{subsec.datasets} (``Datasets'').

\begin{figure*}[t]
\centering
\begin{tabular}{|c|c|c|}
\hline
  \includegraphics[width=2in]{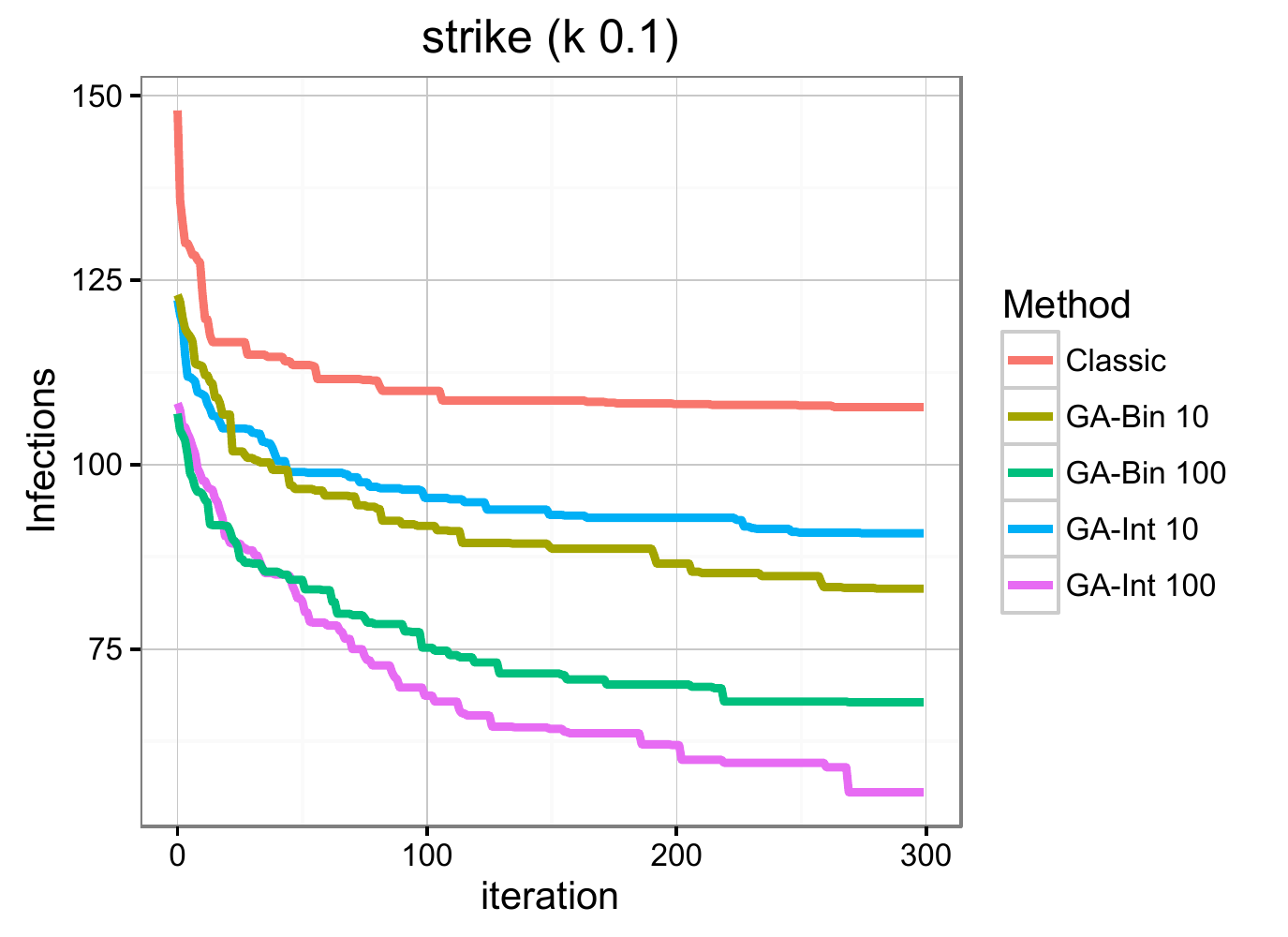} &
  \includegraphics[width=2in]{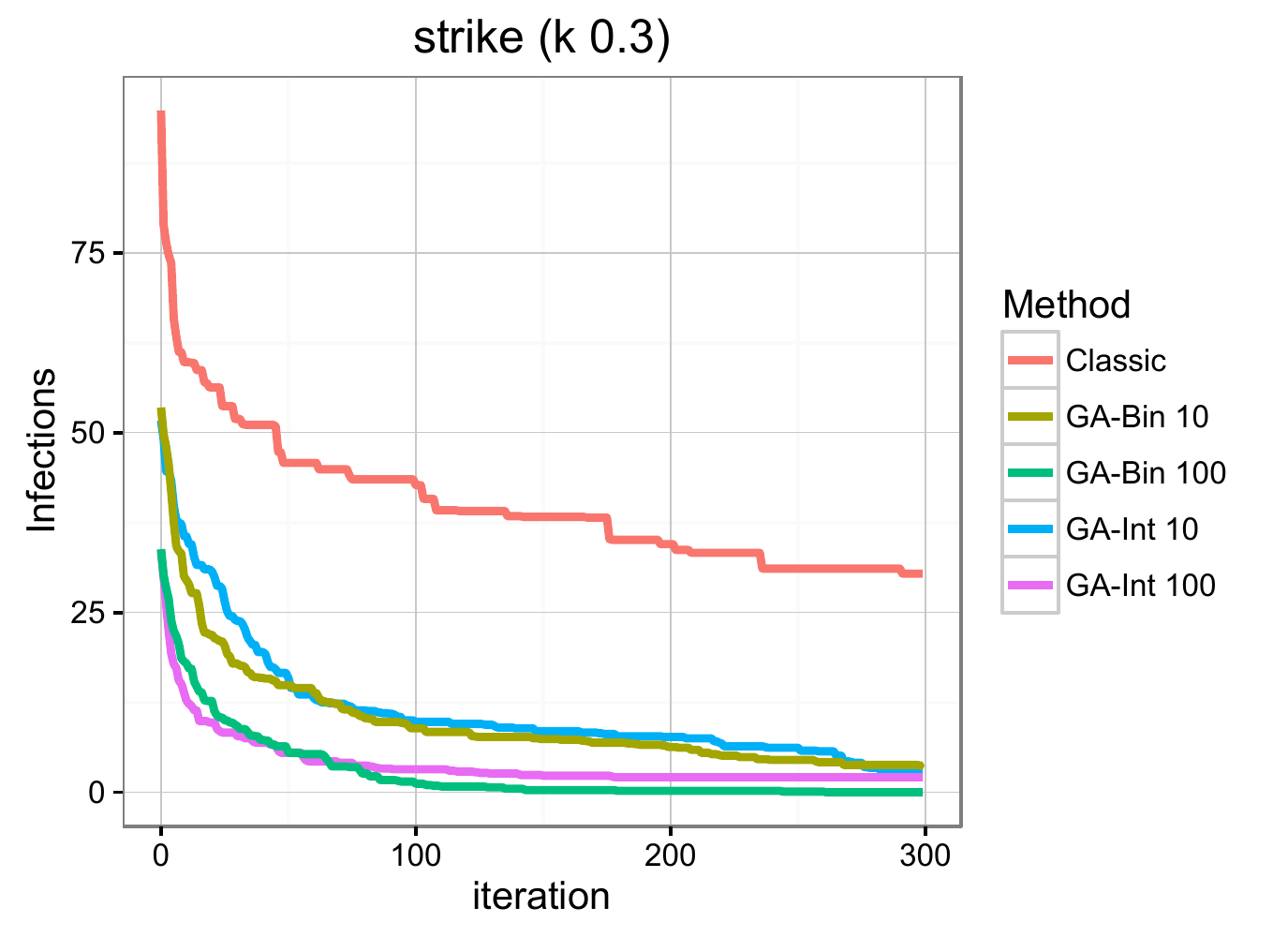} &
  \includegraphics[width=2in]{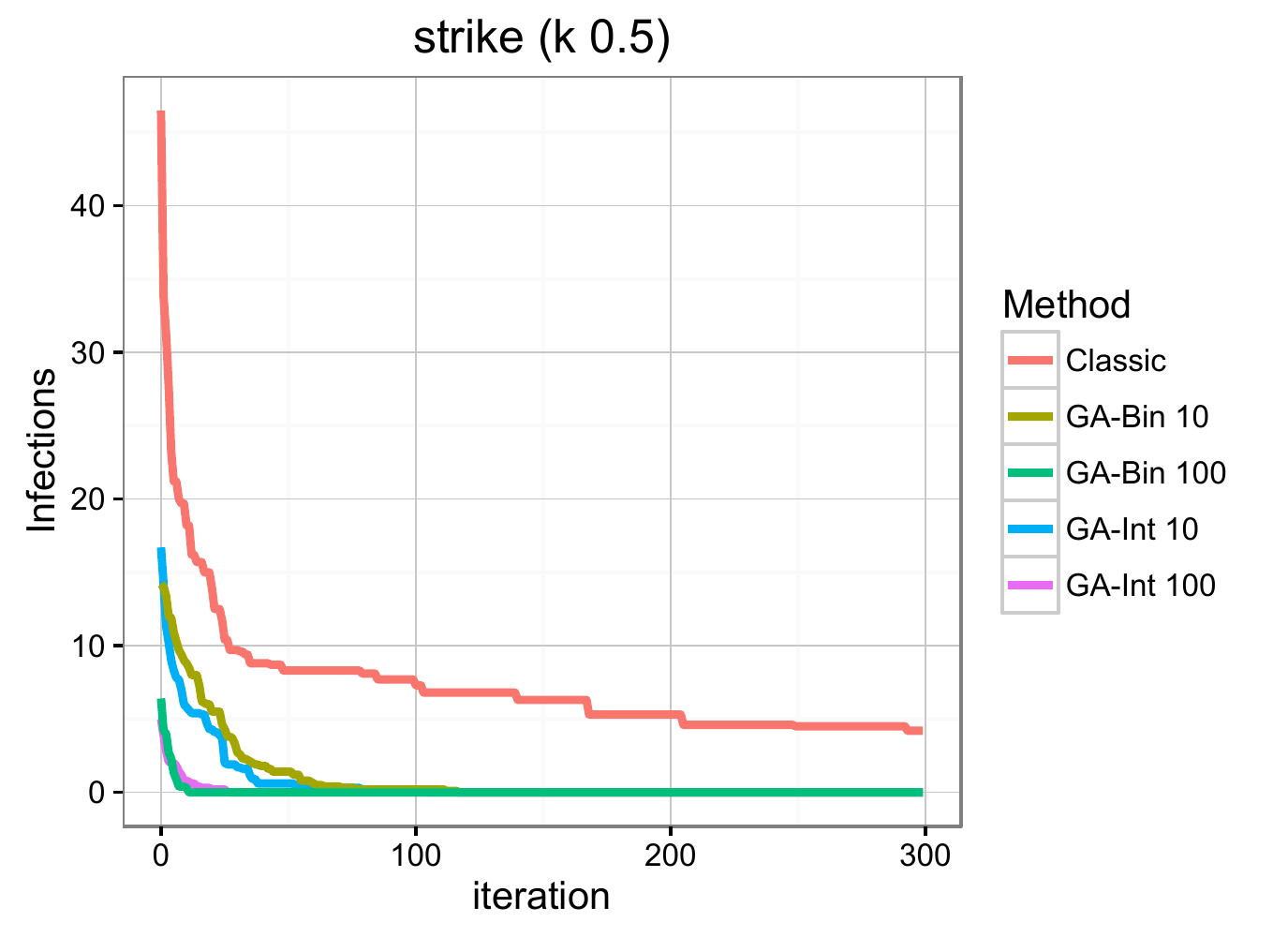} \\
  \hline
  \includegraphics[width=2in]{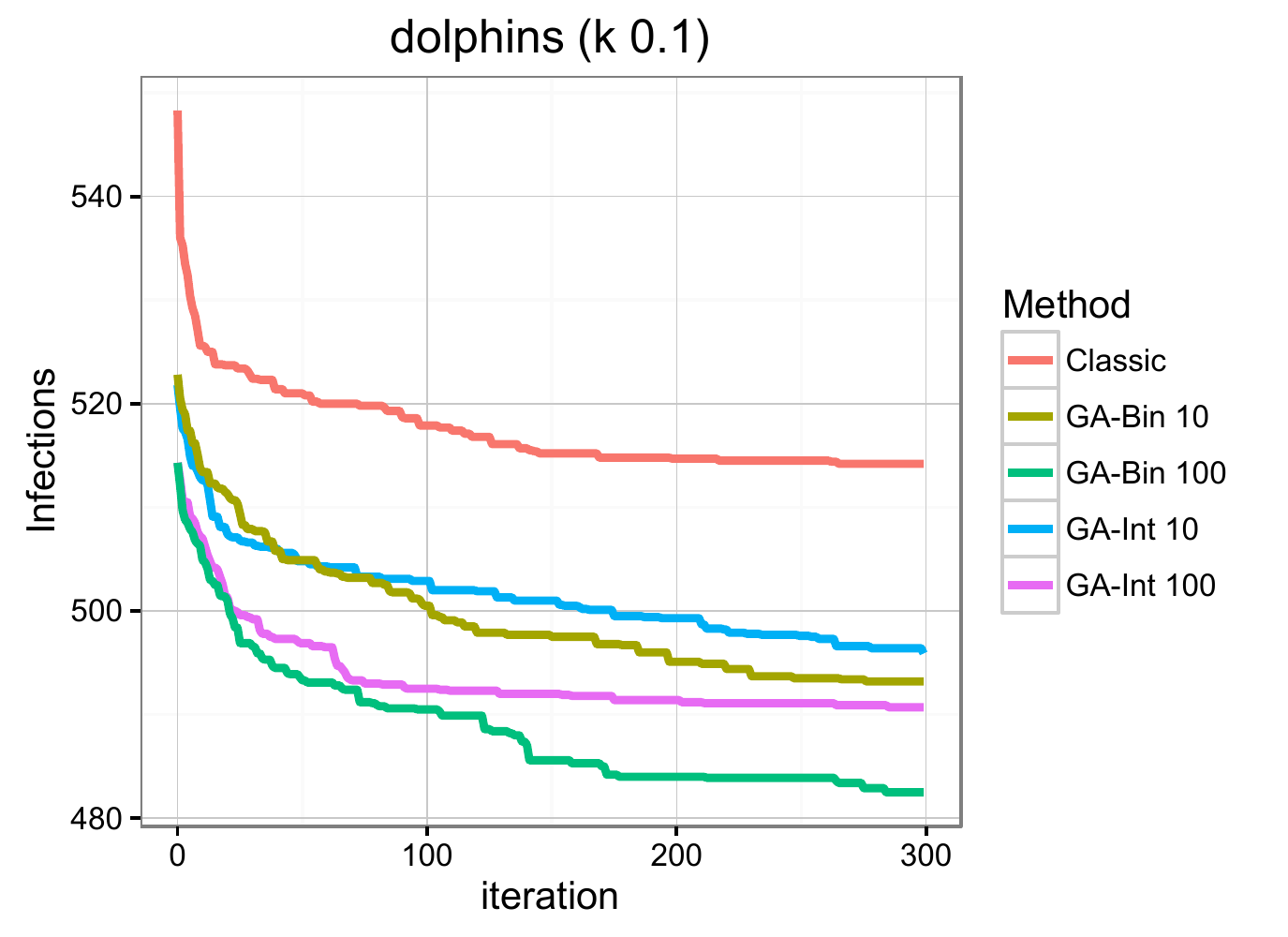} &
  \includegraphics[width=2in]{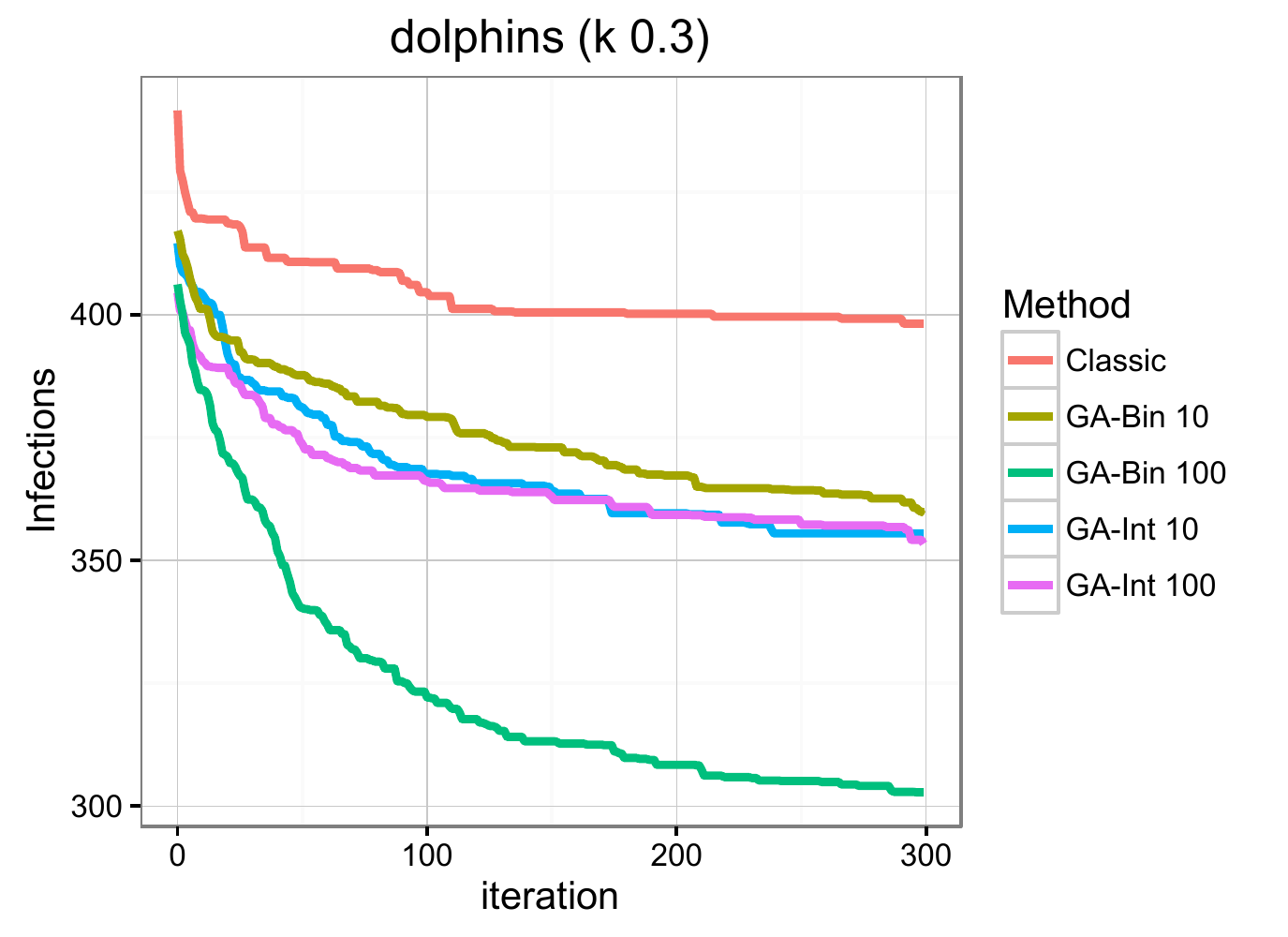} &
  \includegraphics[width=2in]{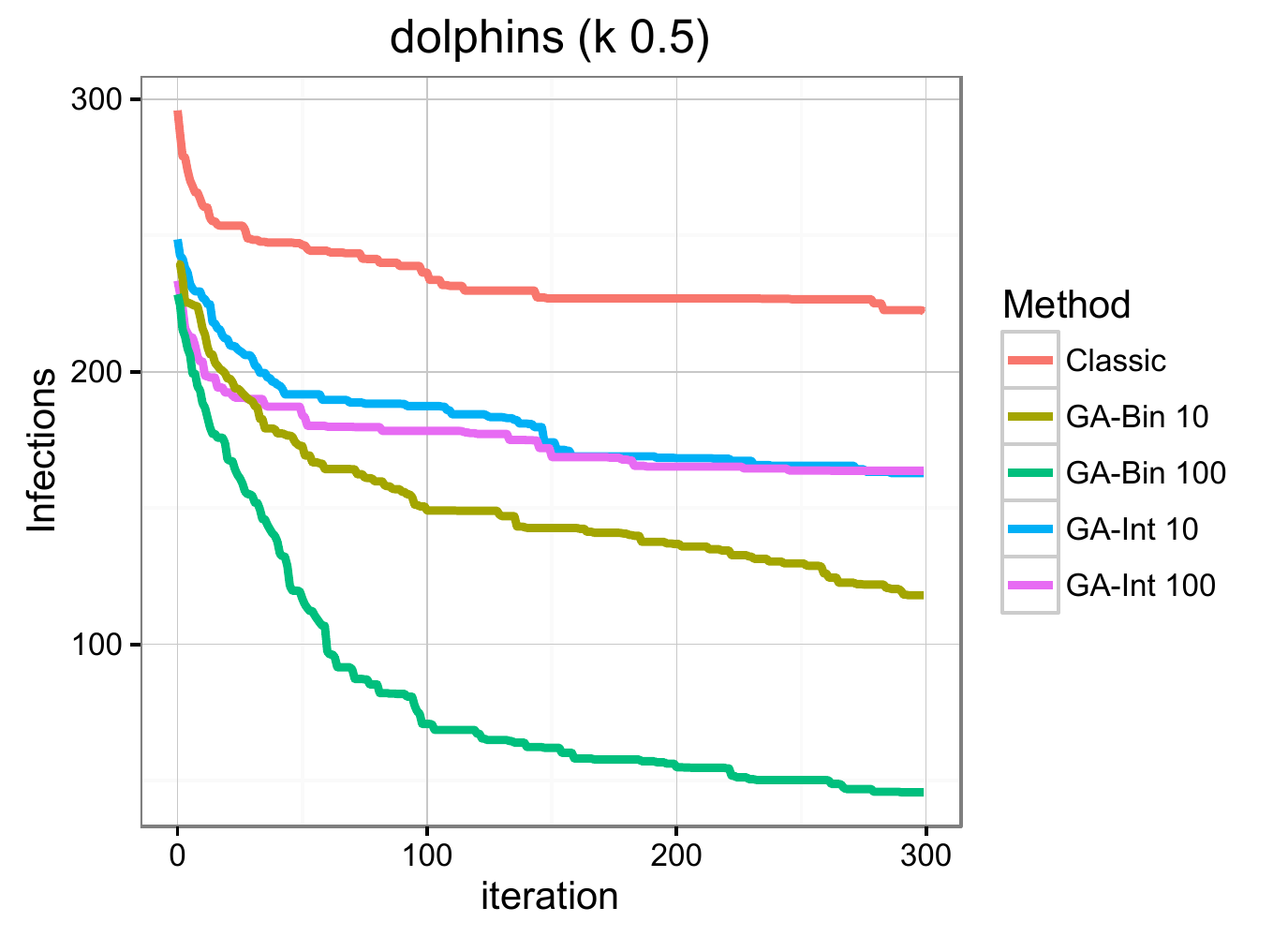} \\
  \hline
  \includegraphics[width=2in]{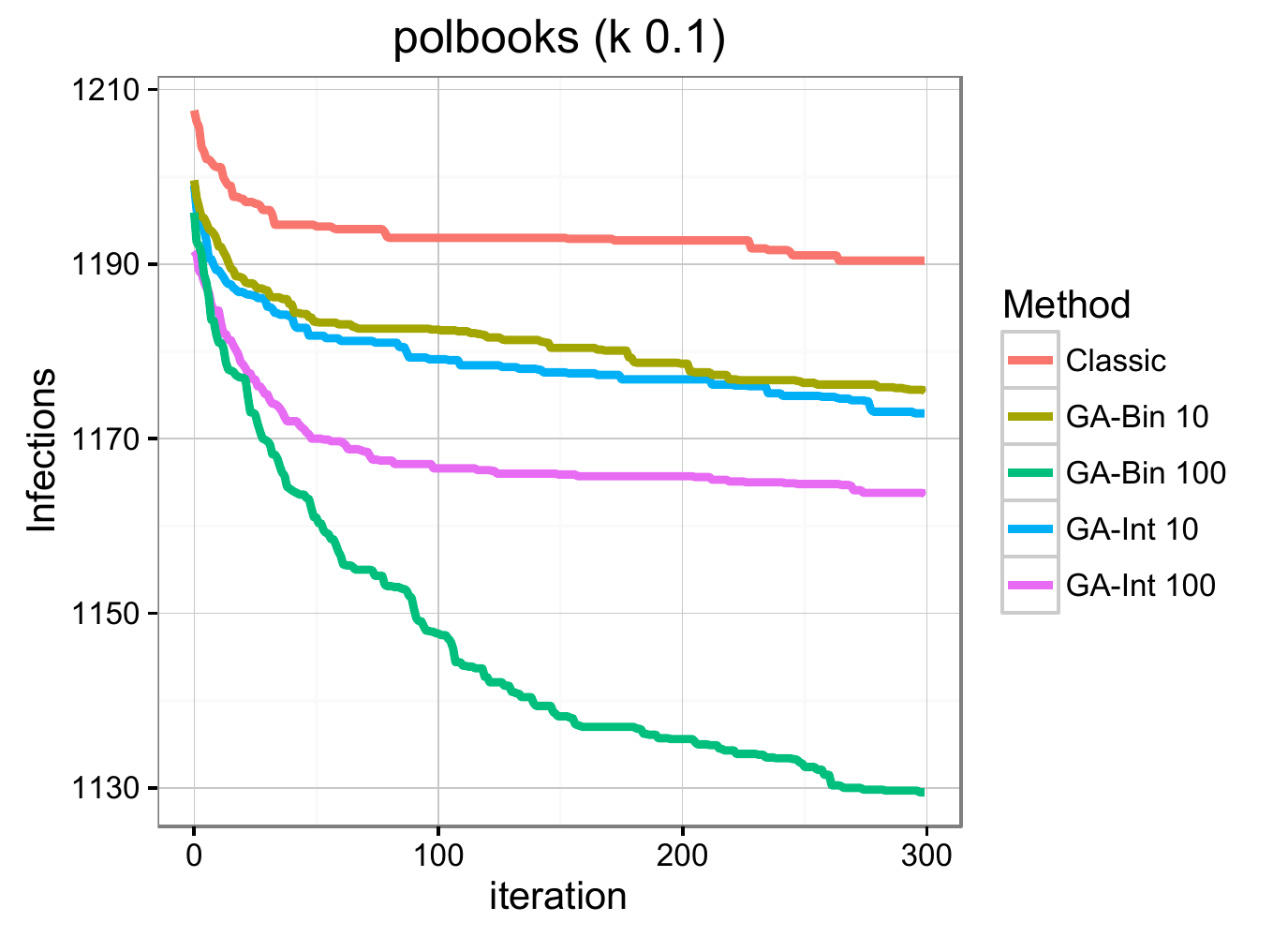} &
  \includegraphics[width=2in]{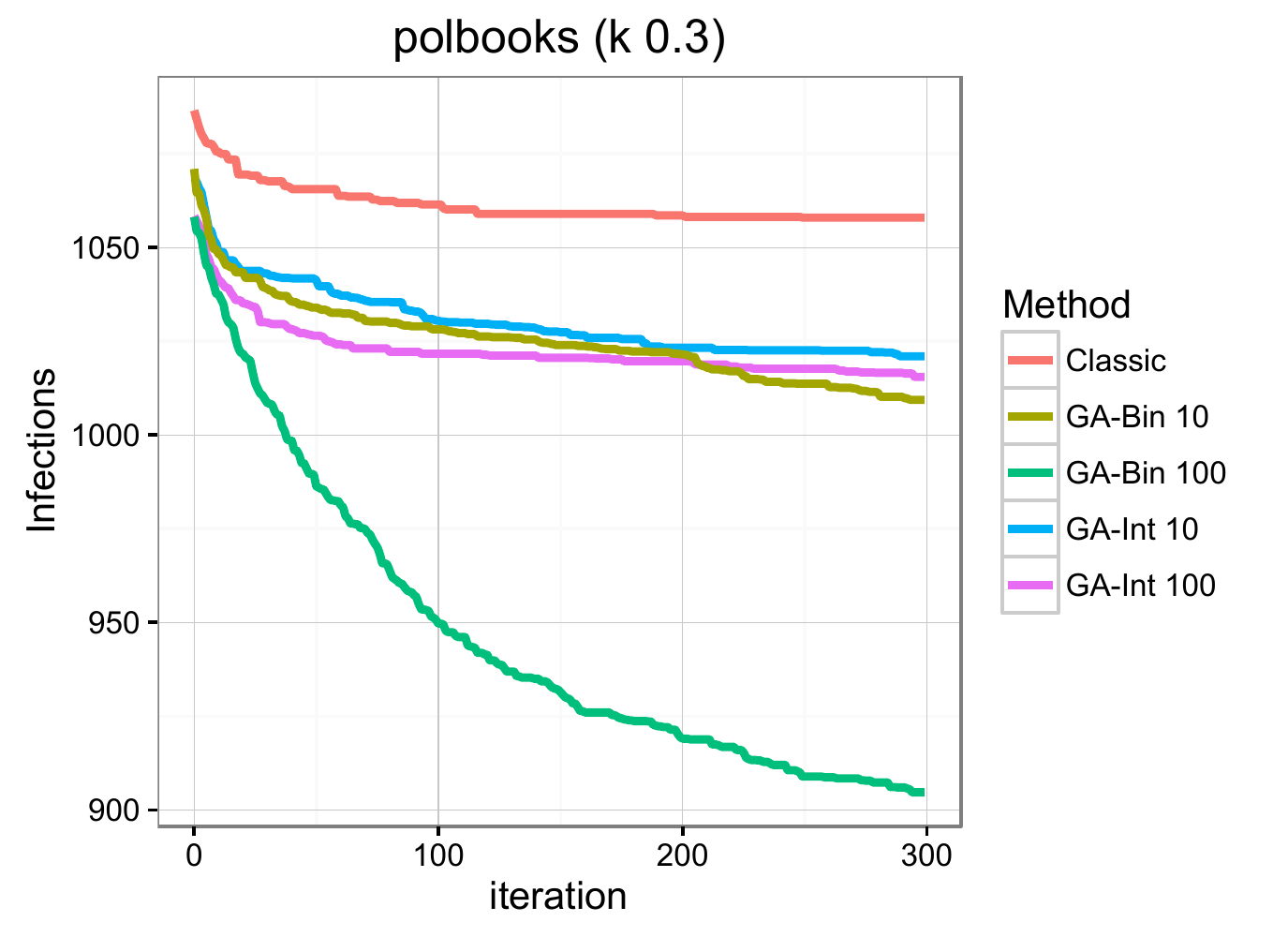} &
  \includegraphics[width=2in]{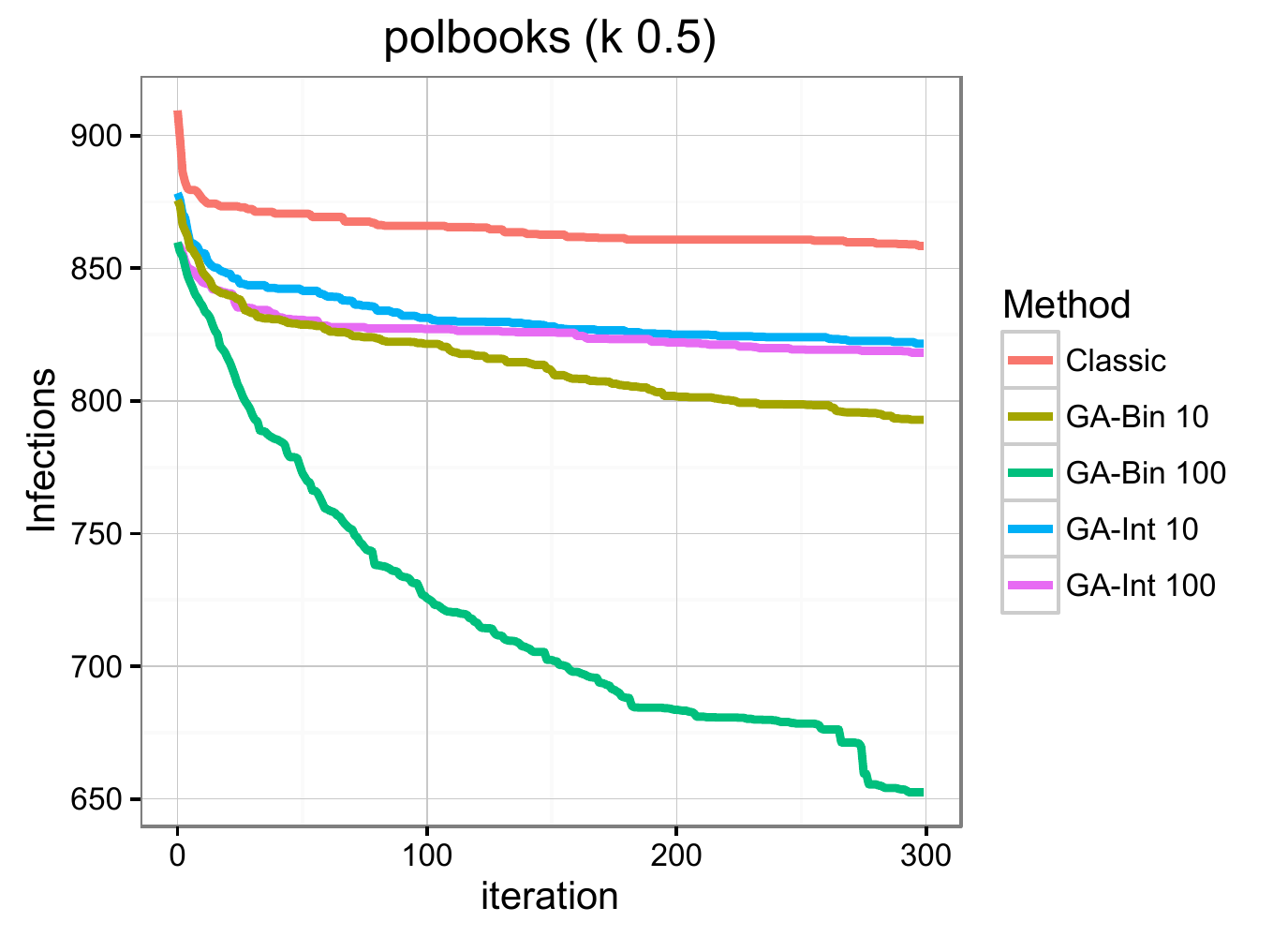} \\
  \hline
  \end{tabular}
  \caption{Average number of infected individuals for Strike, Dolphins, and Polbooks.}
  \label{fig.evolution1}
  \end{figure*}

  \begin{figure*}[t]
  \centering
  \begin{tabular}{|c|c|c|}
  \hline
  \includegraphics[width=2in]{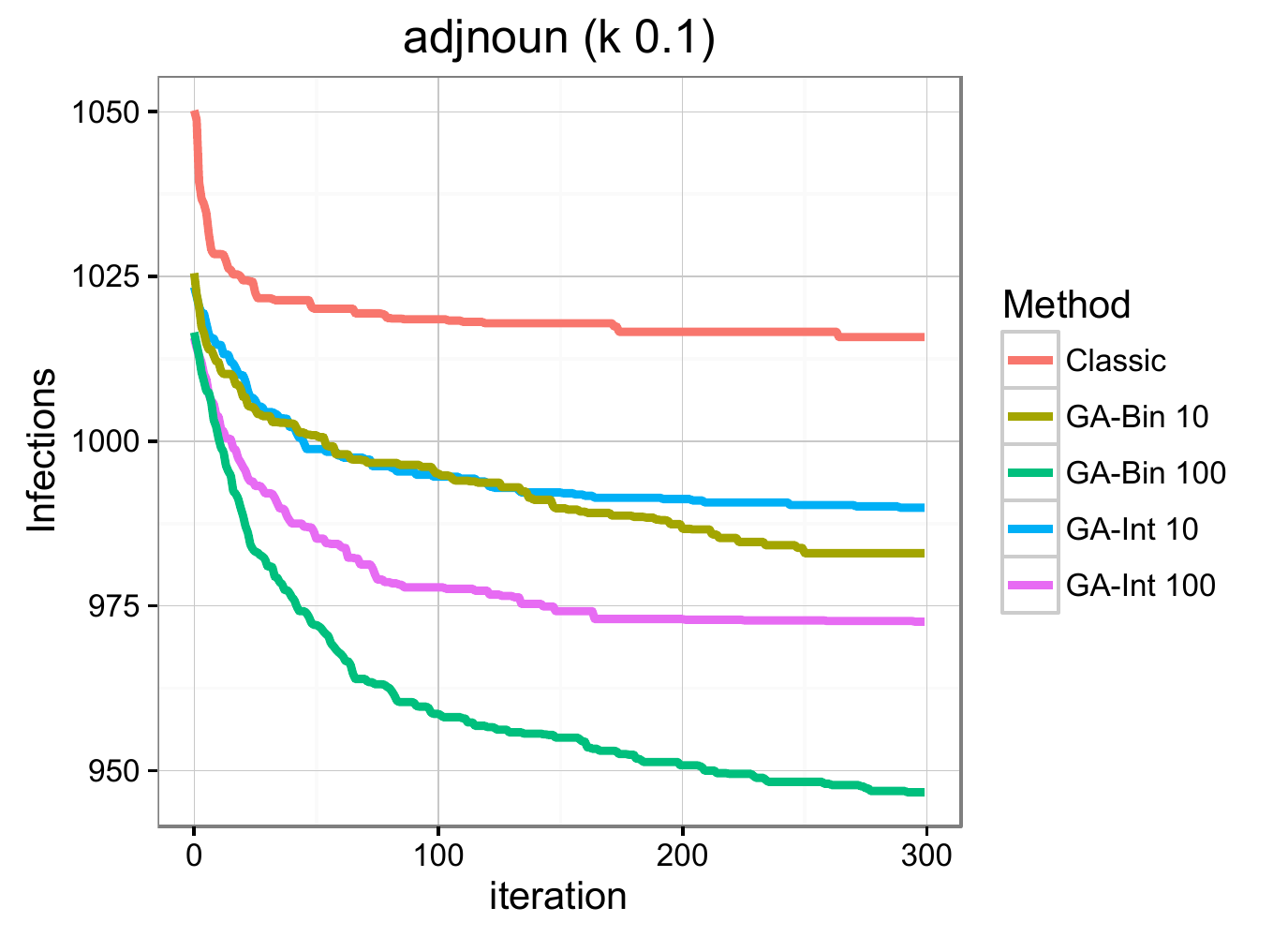} &
  \includegraphics[width=2in]{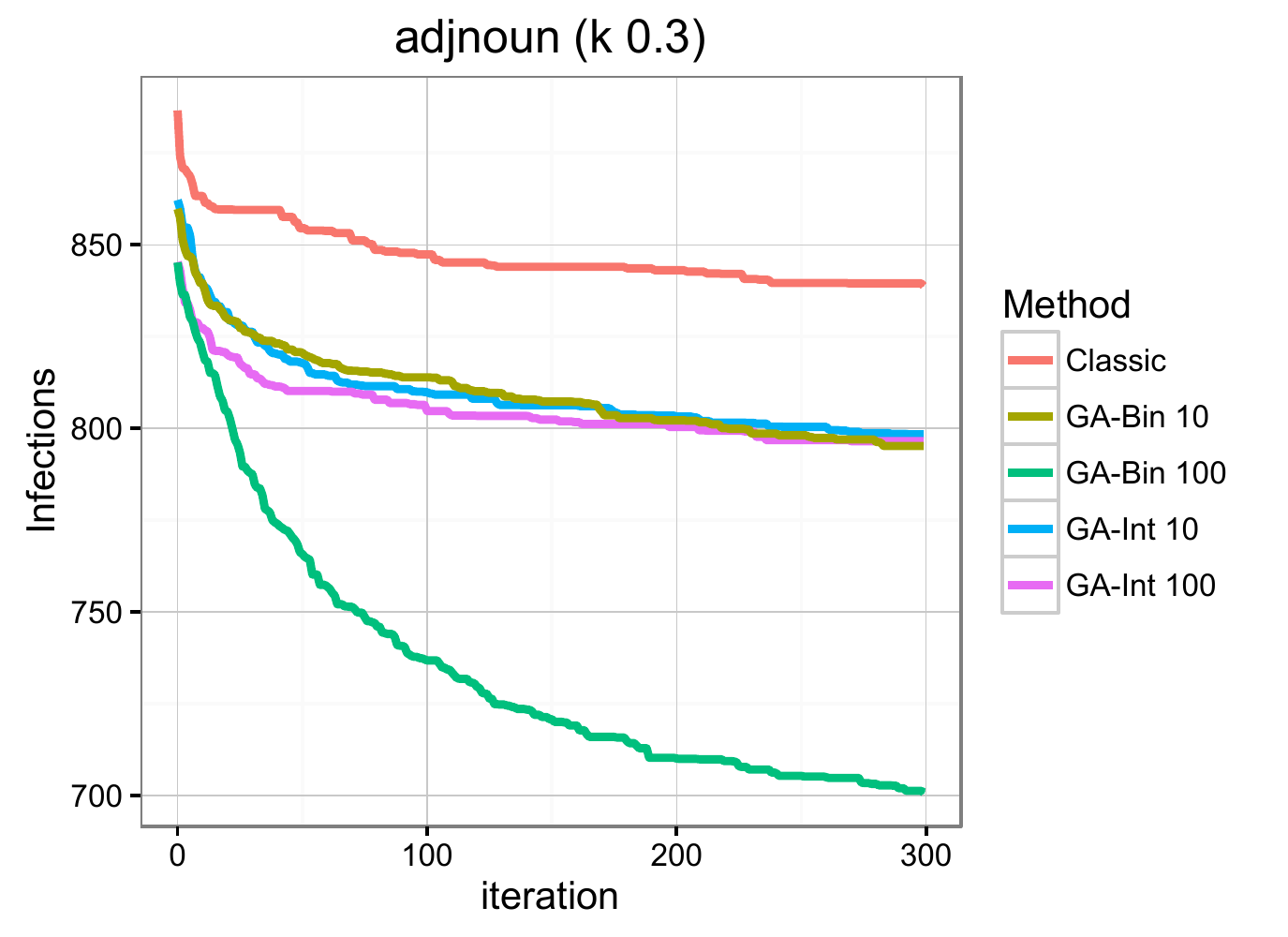} &
  \includegraphics[width=2in]{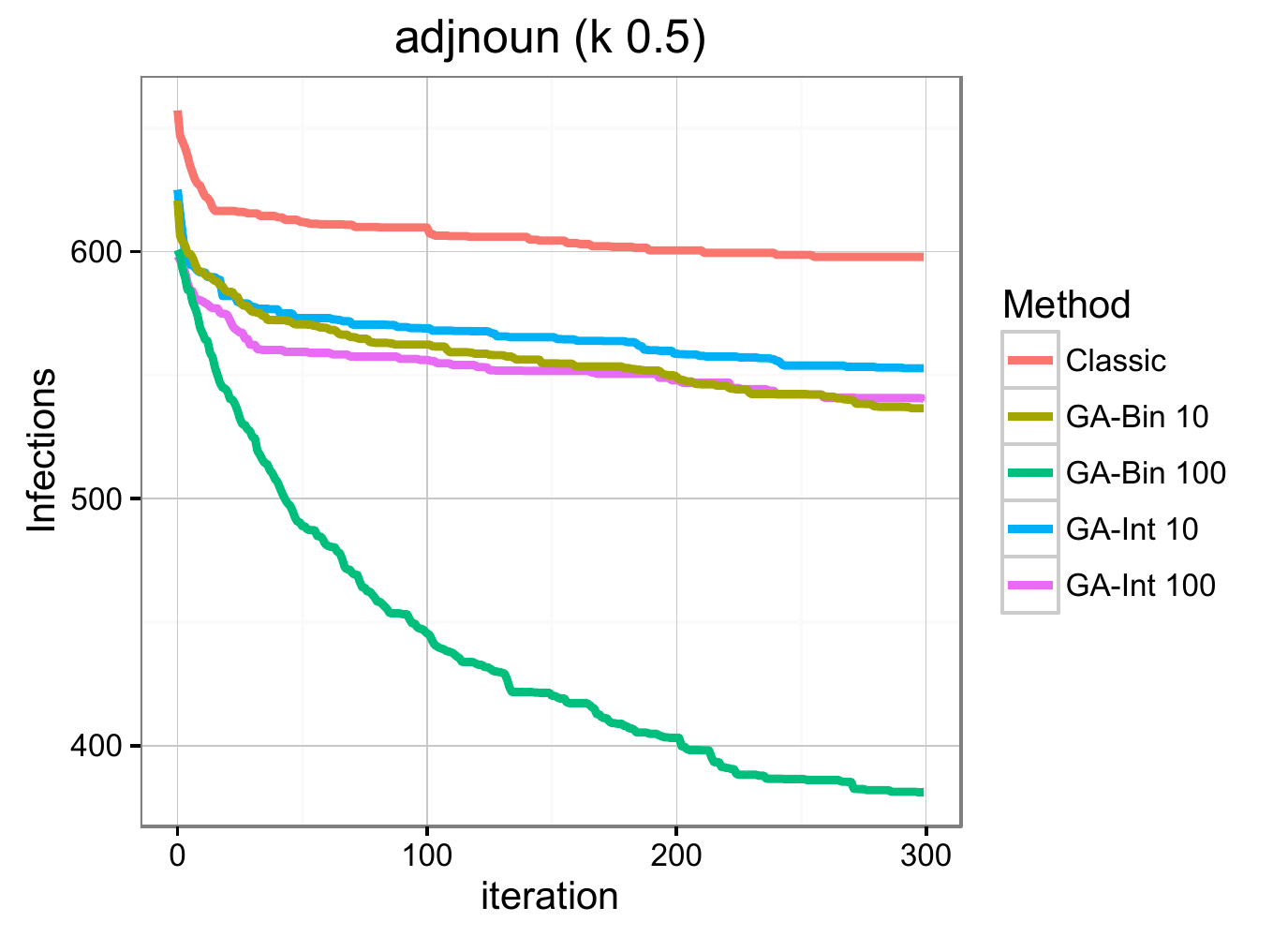} \\
  \hline
  \includegraphics[width=2in]{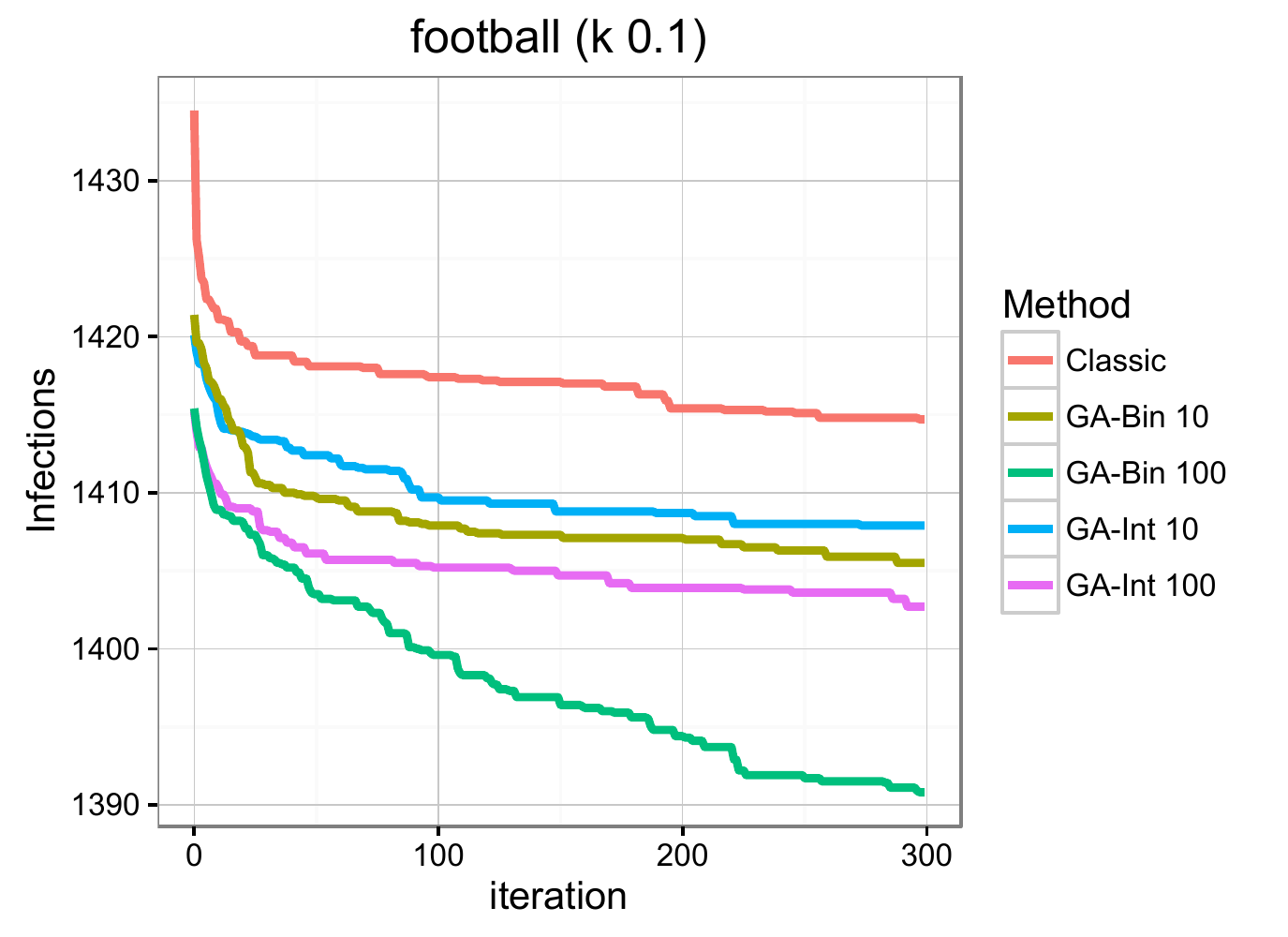} &
  \includegraphics[width=2in]{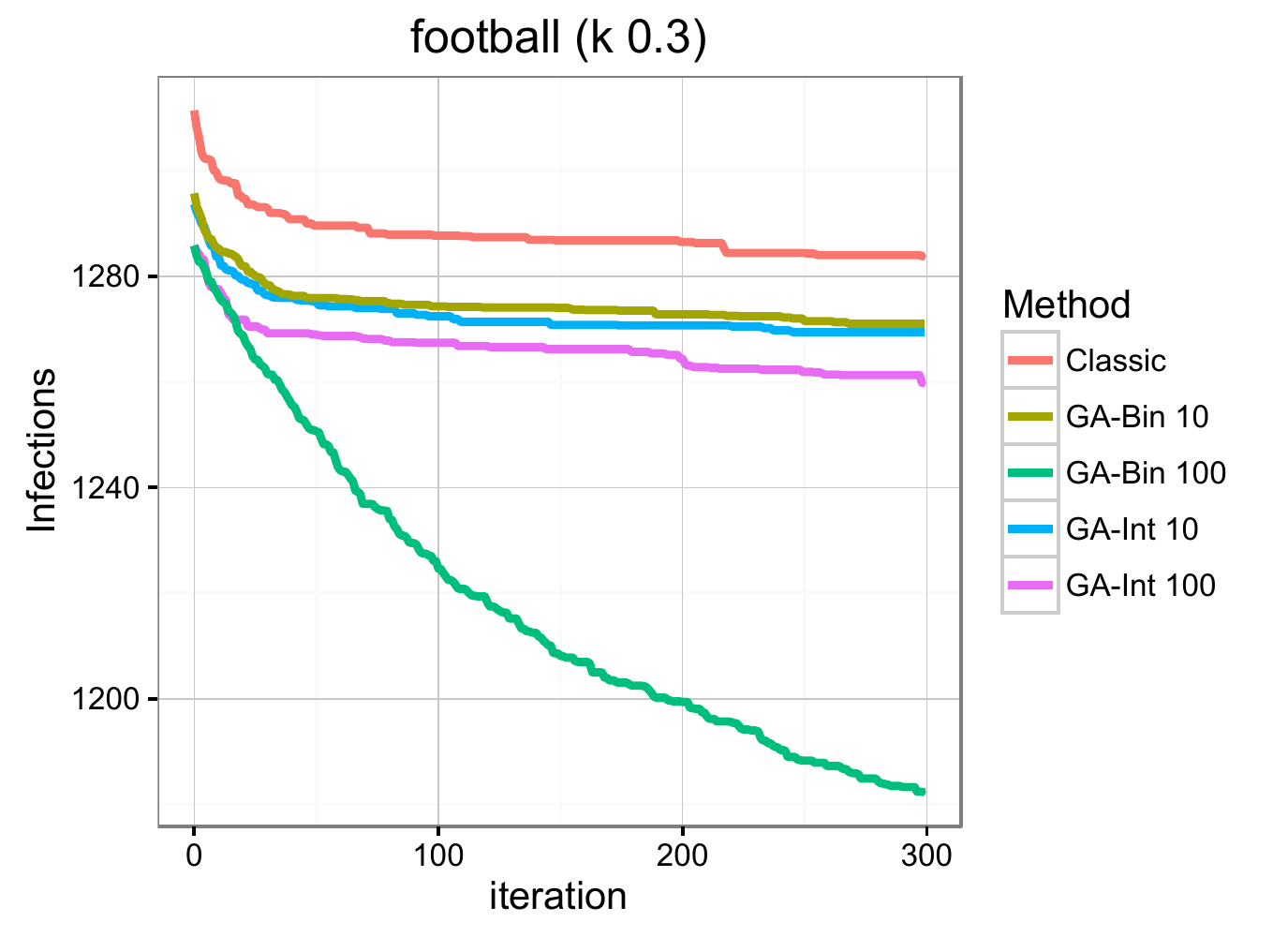} &
  \includegraphics[width=2in]{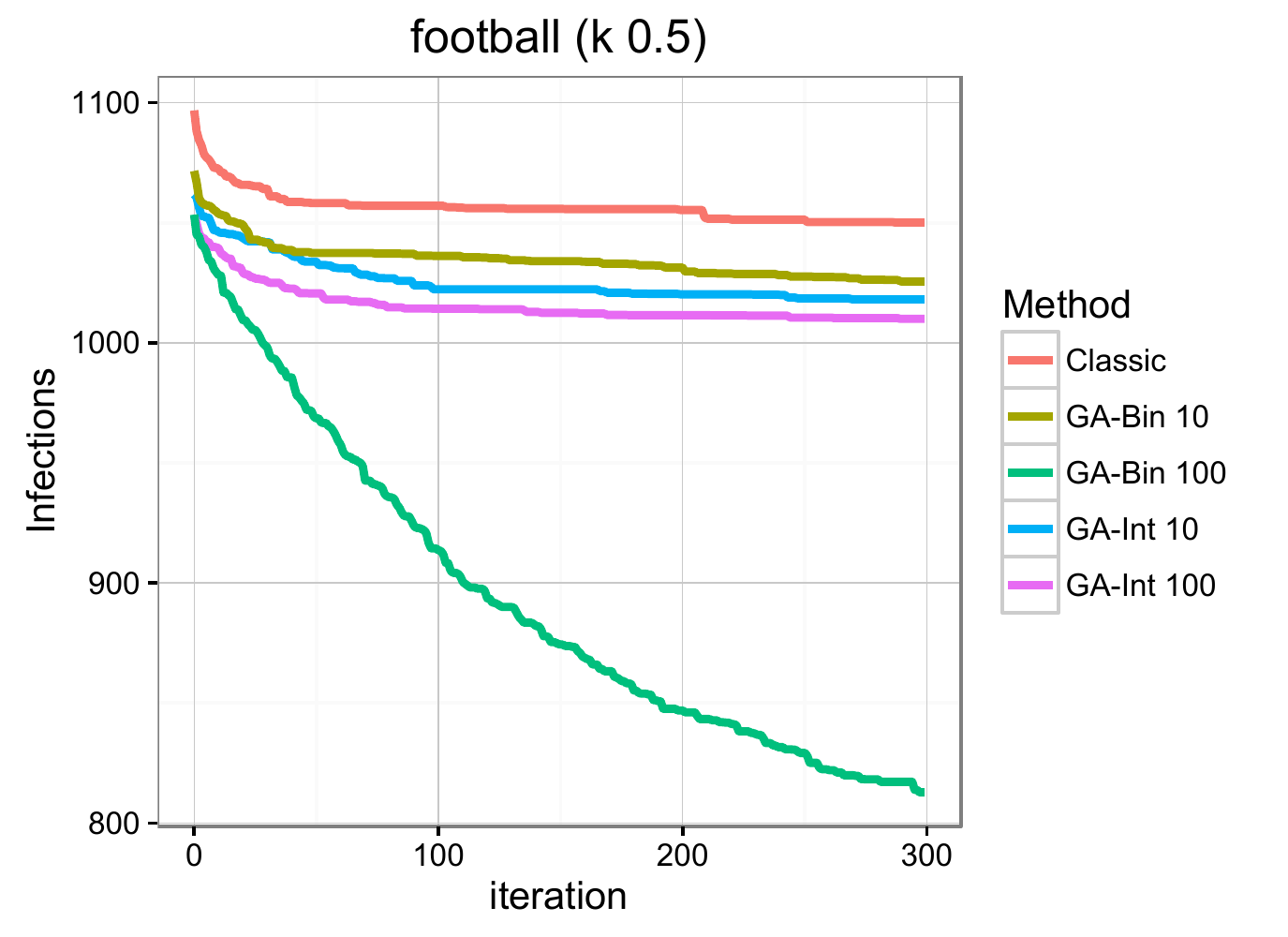} \\
  \hline
  \includegraphics[width=2in]{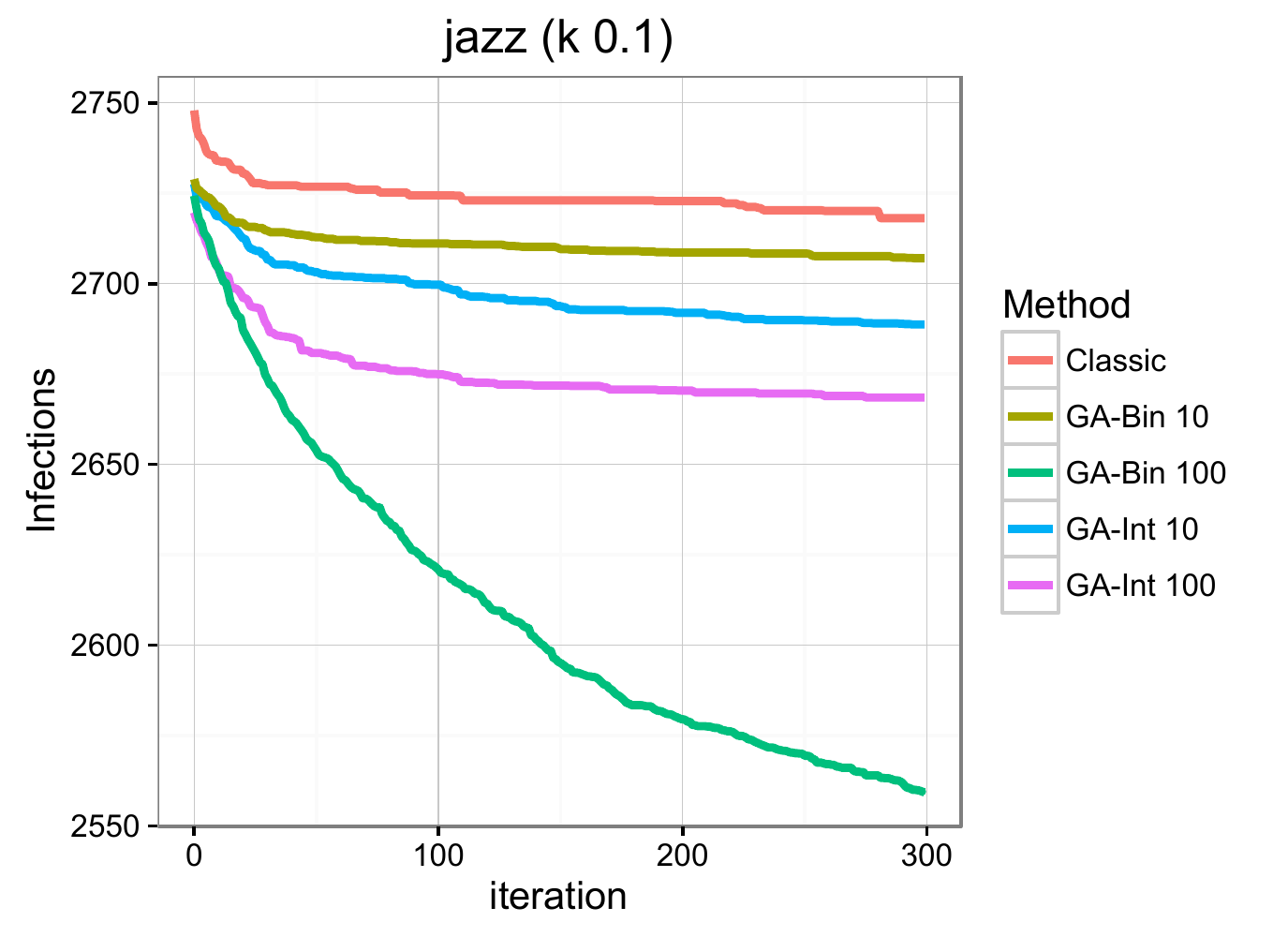} &
  \includegraphics[width=2in]{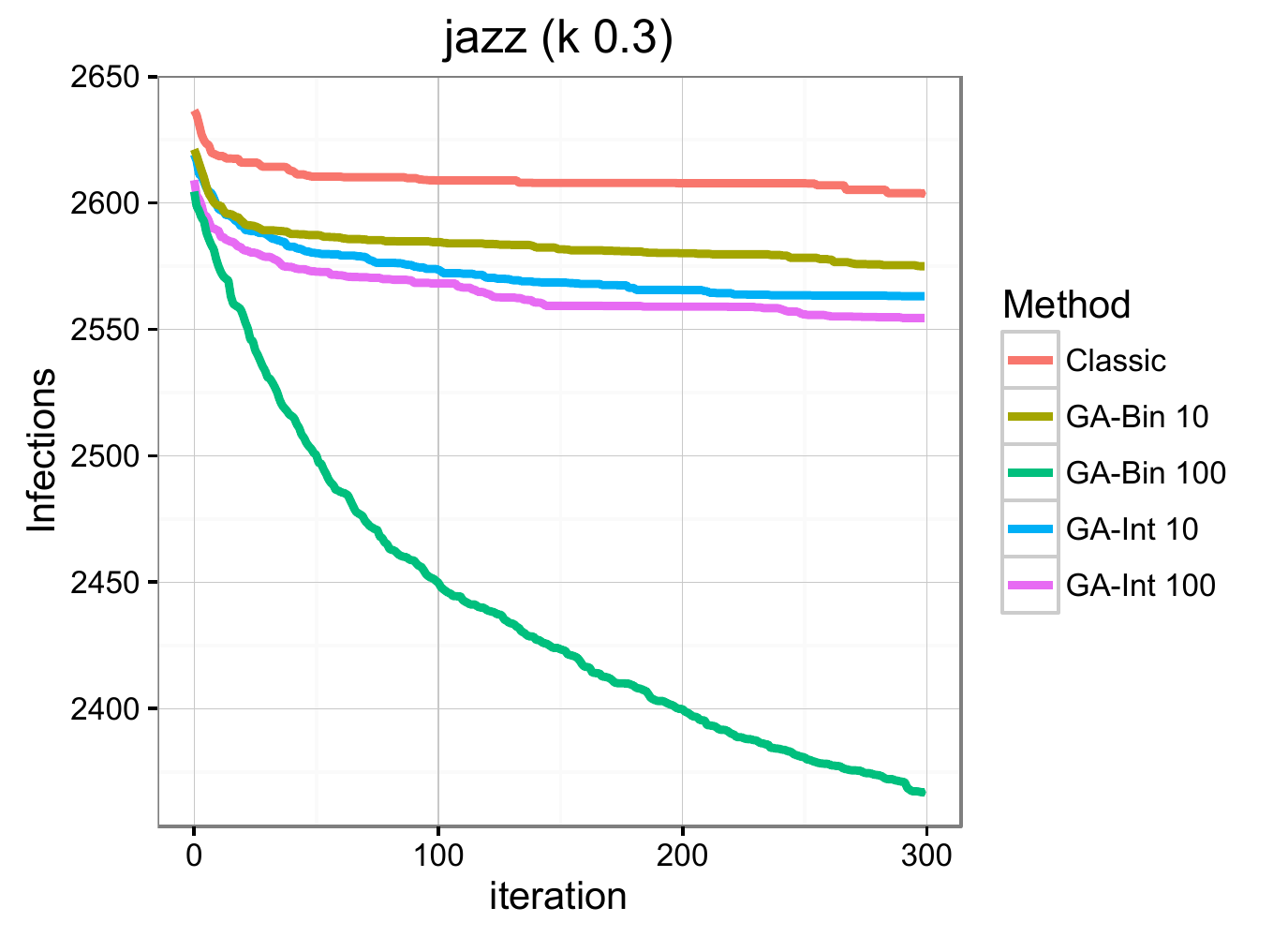} &
  \includegraphics[width=2in]{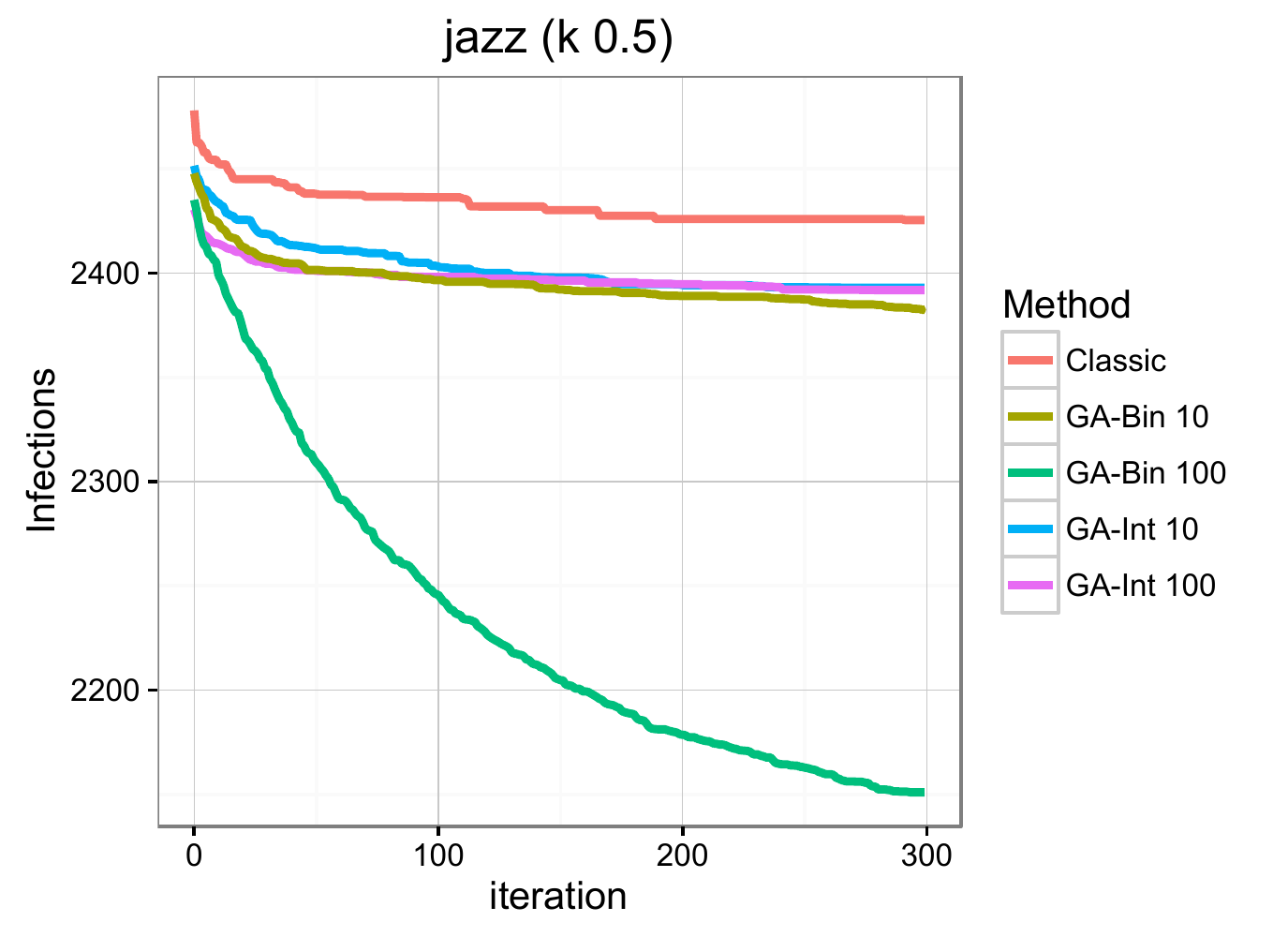} \\
\hline
\end{tabular}
\caption{Average number of infected individuals for Adjnoun, Football, and Jazz.}
\label{fig.evolution2}
\end{figure*}

Tables \ref{tab.alg.gt.k0.1}, \ref{tab.alg.gt.k0.3}, and \ref{tab.alg.gt.k0.5} show the average of the best solutions obtained in each iteration/generation for each $k \in \{0.1\cdot |E|, 0.3\cdot |E|, 0.5\cdot |E|\}$ where E is the set of connections of the network. The standard error is in the right-side of each average. Classic is the heuristic of \cite{Santiago2016} (section \ref{sec.heuristic}), and GA-Int and GA-Bin are the genetic algorithms proposed in this paper with integer and binary chromosomes respectively (Section \ref{sec.ga}). The number on the right-side of the label of genetic algorithm is the number of individuals in the population.

Our genetic algorithms surpassed the Classic heuristic for all results. The genetic algorithm with binary chromosomes and 100 individuals (GA-Bin 100) obtained the best results. Between the genetic algorithms that used 10 individuals in their population, GA-Bin 10 obtained the best results. Two important parameters arose in this comparison. The size of the population and the genetic representation of the solution domain are important factors to obtain solutions with minimal infection spreading.

\begin{table}[h]
\renewcommand{\arraystretch}{1.0}
\caption{Average and standard error from the results of the number of infected individuals by each heuristic when $k=0.1\cdot |E|$.}
\label{tab.alg.gt.k0.1}
\centering
\resizebox{3.2in}{!}{
\begin{tabular}{lrrrrr}
\toprule
\multicolumn{1}{l}{ \textbf{Instance} } 	  & \multicolumn{1}{c}{\textbf{Classic}} 	  & \multicolumn{1}{c}{\textbf{GA-Int 10}} 	  & \multicolumn{1}{c}{\textbf{GA-Int 100}} 	  & \multicolumn{1}{c}{\textbf{GA-Bin 10}} 	  & \multicolumn{1}{c}{\textbf{GA-Bin 100}} 	 \\
\cmidrule(lr){1-1} \cmidrule(lr){2-2}\cmidrule(lr){3-3}\cmidrule(lr){4-4}\cmidrule(lr){5-5}\cmidrule(lr){6-6}strike 	 &107.8$\pm$1.84 	 &90.7$\pm$3.006 	 &\textbf{55.6$\pm$4.15} 	 &83.2$\pm$5.08 	 &67.8$\pm$3.89 	 \\
karate 	 &294.9$\pm$0.81 	 &278.3$\pm$4.08 	 &260.4$\pm$1.07 	 &278.6$\pm$3.57 	 &\textbf{259.2$\pm$0.86} 	 \\
korea1 	 &184.7$\pm$1.61 	 &167.1$\pm$1.91 	 &147.1$\pm$15.53 	 &151.6$\pm$15.92 	 &\textbf{146.7$\pm$15.48} 	 \\
korea2 	 &246.4$\pm$2.58 	 &220.0$\pm$2.32 	 &193.7$\pm$20.01 	 &220.0$\pm$2.98 	 &\textbf{190.4$\pm$19.98} 	 \\
sawmill 	 &181.5$\pm$2.19 	 &155.9$\pm$5.75 	 &137.8$\pm$4.72 	 &155.4$\pm$5.27 	 &\textbf{135.7$\pm$9.7} 	 \\
dolphins 	 &514.2$\pm$1.48 	 &495.9$\pm$1.41 	 &490.7$\pm$1.59 	 &493.2$\pm$2.42 	 &\textbf{482.5$\pm$1.65} 	 \\
football 	 &1414.7$\pm$0.6 	 &1407.9$\pm$0.92 	 &1402.7$\pm$0.72 	 &1405.5$\pm$0.45 	 &\textbf{1390.8$\pm$1.38} 	 \\
adjnoun 	 &1015.8$\pm$2.07 	 &989.9$\pm$1.67 	 &972.6$\pm$2.18 	 &983.0$\pm$3.78 	 &\textbf{946.7$\pm$1.25} 	 \\
polbooks 	 &1190.4$\pm$2.3 	 &1172.9$\pm$3.84 	 &1163.7$\pm$2.49 	 &1175.5$\pm$3.02 	 &\textbf{1129.5$\pm$1.91} 	 \\
jazz 	 &2718.1$\pm$1.82 	 &2688.7$\pm$2.23 	 &2668.5$\pm$1.62 	 &2707.0$\pm$3.06 	 &\textbf{2559.3$\pm$2.6} 	 \\
\bottomrule
\end{tabular}}
\end{table}

\begin{table}[h]
\renewcommand{\arraystretch}{1.0}
\caption{Average and standard error from the results of the number of infected individuals by each heuristic when $k=0.3\cdot |E|$.}
\label{tab.alg.gt.k0.3}
\centering
\resizebox{3.2in}{!}{
\begin{tabular}{lrrrrr}
\toprule
\multicolumn{1}{l}{ \textbf{Instance} } 	  & \multicolumn{1}{c}{\textbf{Classic}} 	  & \multicolumn{1}{c}{\textbf{GA-Int 10}} 	  & \multicolumn{1}{c}{\textbf{GA-Int 100}} 	  & \multicolumn{1}{c}{\textbf{GA-Bin 10}} 	  & \multicolumn{1}{c}{\textbf{GA-Bin 100}} 	 \\
\cmidrule(lr){1-1} \cmidrule(lr){2-2}\cmidrule(lr){3-3}\cmidrule(lr){4-4}\cmidrule(lr){5-5}\cmidrule(lr){6-6}strike 	 &30.4$\pm$3.98 	 &3.1$\pm$0.78 	 &2.1$\pm$0.65 	 &3.7$\pm$1.52 	 &\textbf{.0$\pm$.0} 	 \\
karate 	 &230.0$\pm$0.63 	 &203.8$\pm$2.55 	 &201.6$\pm$1.78 	 &192.1$\pm$1.99 	 &\textbf{184.0$\pm$2.04} 	 \\
korea1 	 &114.5$\pm$11.78 	 &24.0$\pm$12.66 	 &\textbf{.0$\pm$.0} 	 &20.7$\pm$12.62 	 &\textbf{.0$\pm$.0} 	 \\
korea2 	 &170.0$\pm$17.41 	 &142.8$\pm$15.16 	 &147.0$\pm$15.55 	 &143.6$\pm$15.44 	 &\textbf{133.8$\pm$14.18} 	 \\
sawmill 	 &80.6$\pm$6.83 	 &14.4$\pm$5.78 	 &9.6$\pm$3.48 	 &14.3$\pm$4.18 	 &\textbf{5.2$\pm$2.23} 	 \\
dolphins 	 &398.2$\pm$2.008 	 &355.5$\pm$3.77 	 &353.4$\pm$3.37 	 &359.8$\pm$5.18 	 &\textbf{302.8$\pm$2.69} 	 \\
football 	 &1283.8$\pm$1.34 	 &1269.4$\pm$0.76 	 &1259.8$\pm$1.57 	 &1271.0$\pm$4.18 	 &\textbf{1182.3$\pm$3.74} 	 \\
adjnoun 	 &838.8$\pm$2.32 	 &798.4$\pm$2.68 	 &796.5$\pm$1.77 	 &795.2$\pm$5.48 	 &\textbf{700.9$\pm$3.6} 	 \\
polbooks 	 &1058.0$\pm$2.15 	 &1021.0$\pm$3.33 	 &1015.5$\pm$1.54 	 &1009.4$\pm$4.75 	 &\textbf{904.7$\pm$3.29} 	 \\
jazz 	 &2603.8$\pm$2.25 	 &2563.1$\pm$1.87 	 &2554.5$\pm$2.28 	 &2575.0$\pm$2.51 	 &\textbf{2366.9$\pm$3.45} 	 \\
\bottomrule
\end{tabular}}
\end{table}

\begin{table}[h]
\renewcommand{\arraystretch}{1.0}
\caption{Average and standard error from the results of the number of infected individuals by each heuristic when $k=0.5\cdot |E|$.}
\label{tab.alg.gt.k0.5}
\centering
\resizebox{3.2in}{!}{
\begin{tabular}{lrrrrr}
\toprule
\multicolumn{1}{l}{ \textbf{Instance} } 	  & \multicolumn{1}{c}{\textbf{Classic}} 	  & \multicolumn{1}{c}{\textbf{GA-Int 10}} 	  & \multicolumn{1}{c}{\textbf{GA-Int 100}} 	  & \multicolumn{1}{c}{\textbf{GA-Bin 10}} 	  & \multicolumn{1}{c}{\textbf{GA-Bin 100}} 	 \\
\cmidrule(lr){1-1} \cmidrule(lr){2-2}\cmidrule(lr){3-3}\cmidrule(lr){4-4}\cmidrule(lr){5-5}\cmidrule(lr){6-6}strike 	 &4.2$\pm$1.07 	 &\textbf{.0$\pm$.0} 	 &\textbf{.0$\pm$.0} 	 &\textbf{.0$\pm$.0} 	 &\textbf{.0$\pm$.0} 	 \\
karate 	 &132.7$\pm$3.65 	 &40.3$\pm$11.52 	 &4.6$\pm$1.4 	 &27.0$\pm$12.35 	 &\textbf{.0$\pm$.0} 	 \\
korea1 	 &4.0$\pm$1.32 	 &\textbf{.0$\pm$.0} 	 &\textbf{.0$\pm$.0} 	 &\textbf{.0$\pm$.0} 	 &\textbf{.0$\pm$.0} 	 \\
korea2 	 &99.0$\pm$10.68 	 &22.1$\pm$9.6 	 &1.5$\pm$0.95 	 &7.8$\pm$3.009 	 &\textbf{.0$\pm$.0} 	 \\
sawmill 	 &11.1$\pm$2.21 	 &1.6$\pm$1.31 	 &.6$\pm$0.37 	 &.9$\pm$0.85 	 &\textbf{.0$\pm$.0} 	 \\
dolphins 	 &222.2$\pm$3.16 	 &162.8$\pm$11.25 	 &163.7$\pm$4.73 	 &118.0$\pm$9.63 	 &\textbf{45.7$\pm$7.26} 	 \\
football 	 &1050.1$\pm$2.91 	 &1018.1$\pm$1.97 	 &1010.0$\pm$1.66 	 &1025.5$\pm$3.88 	 &\textbf{812.8$\pm$10.42} 	 \\
adjnoun 	 &597.9$\pm$2.24 	 &552.8$\pm$3.08 	 &540.4$\pm$3.4 	 &536.6$\pm$9.06 	 &\textbf{381.1$\pm$3.68} 	 \\
polbooks 	 &858.4$\pm$2.32 	 &821.6$\pm$2.06 	 &818.2$\pm$2.53 	 &792.9$\pm$5.92 	 &\textbf{652.5$\pm$25.92} 	 \\
jazz 	 &2425.4$\pm$4.18 	 &2393.0$\pm$2.2 	 &2391.8$\pm$1.44 	 &2382.4$\pm$6.62 	 &\textbf{2151.0$\pm$1.94} 	 \\
\bottomrule
\end{tabular}}
\end{table}

Figures \ref{fig.evolution1} and \ref{fig.evolution2} show the best solution found in each iteration for each tested heuristic in six different instances. In every single iteration, our genetic algorithms surpassed the best solution found in the Classic heuristic. The size of the population and the genetic representation are the most important parameters of the tested genetic algorithms. The size of population equal to 100 and the binary chromosome demonstrated a faster convergence than the other tested methods.

When the network was small enough or when the number of connections to be removed was sufficiently high, our genetic algorithm was able to completely isolate the initially infected individuals, resulting in total containment of the epidemic. Also, higher amounts of removed connections ($k$) tend to result in a larger improvement over the original heuristic in the majority of the networks tested.

\begin{figure*}[ht]
\centering
\begin{tabular}{|c|c|c|}
\hline
\includegraphics[width=2in]{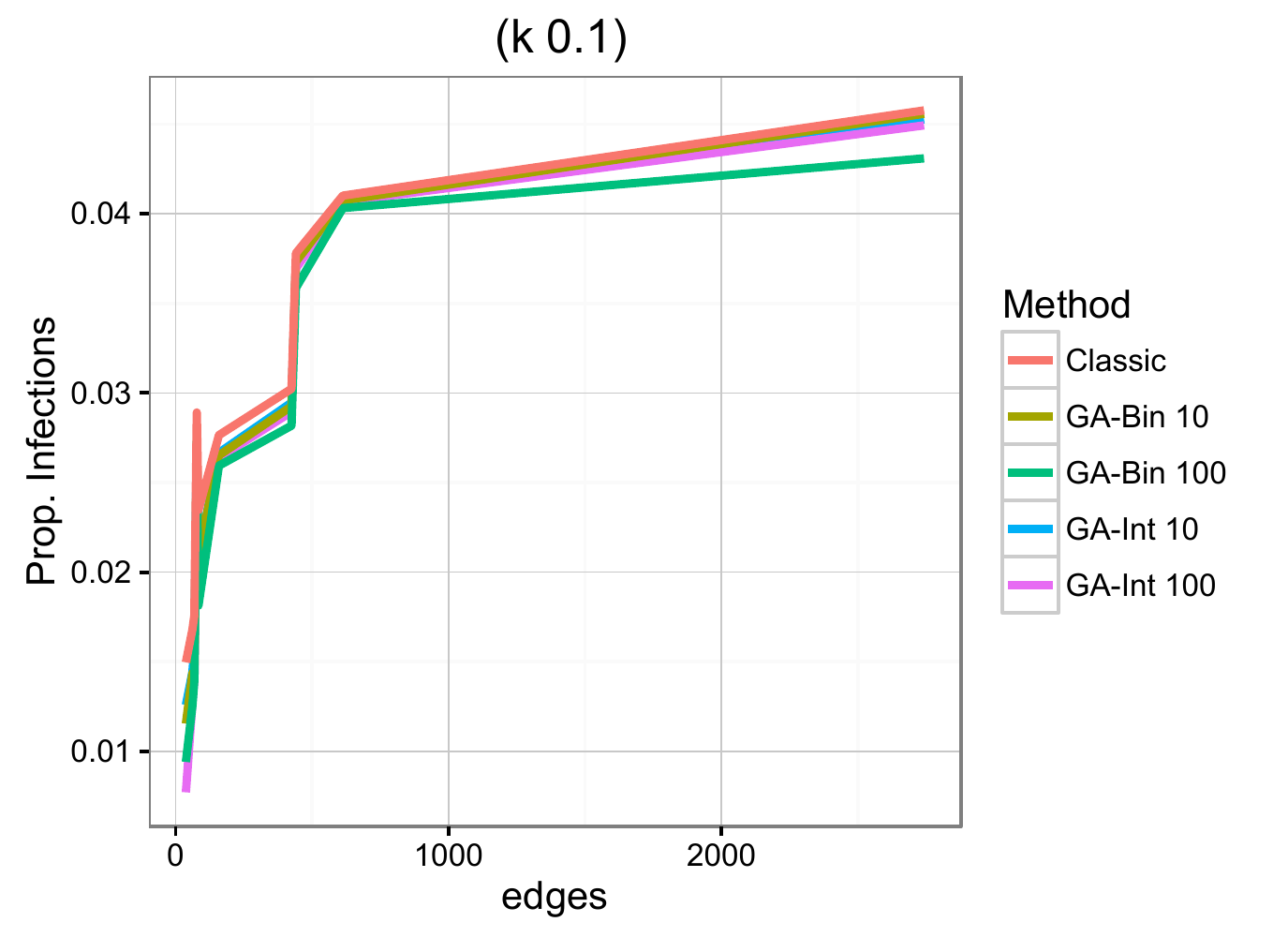} &
\includegraphics[width=2in]{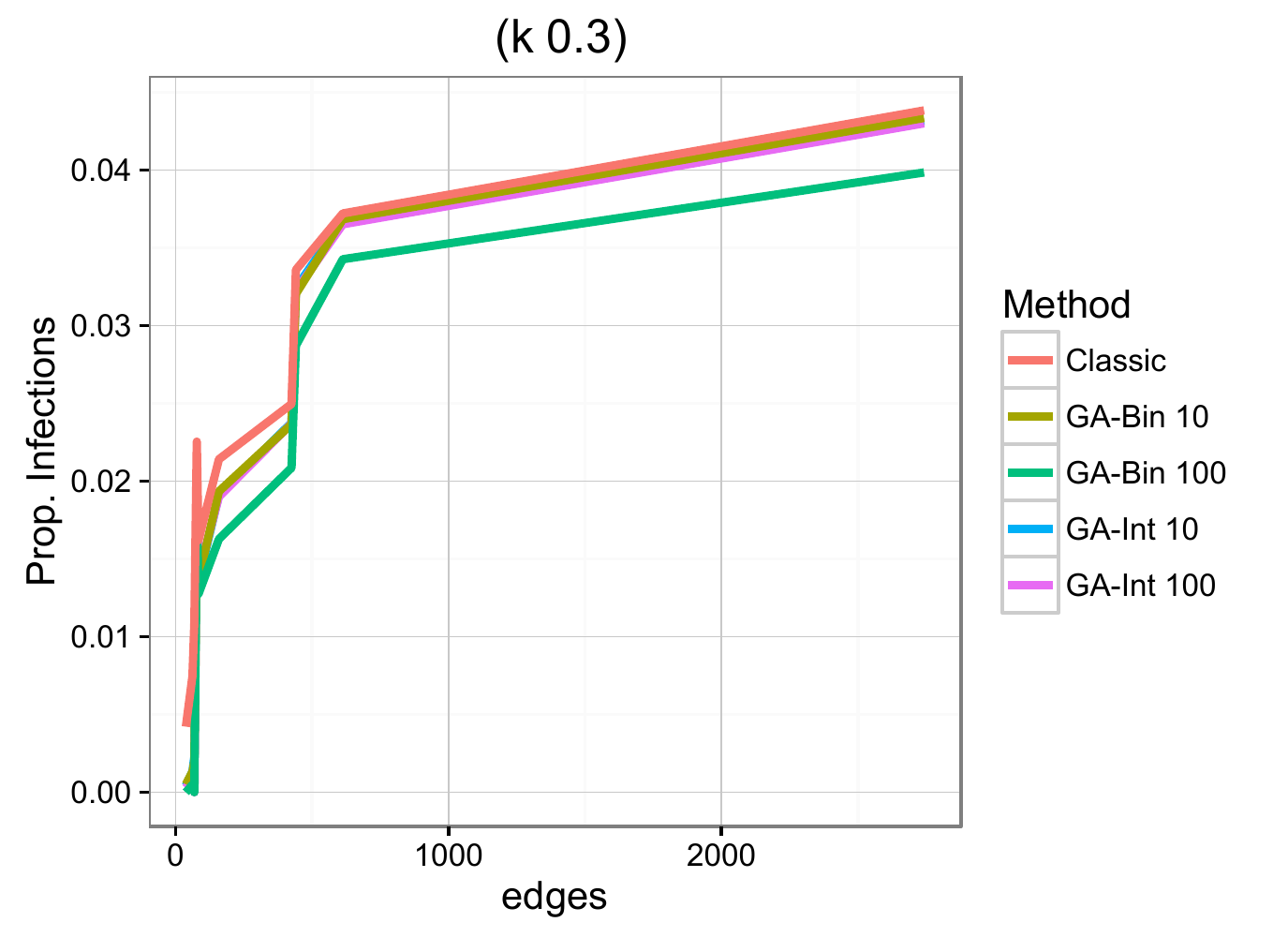} &
\includegraphics[width=2in]{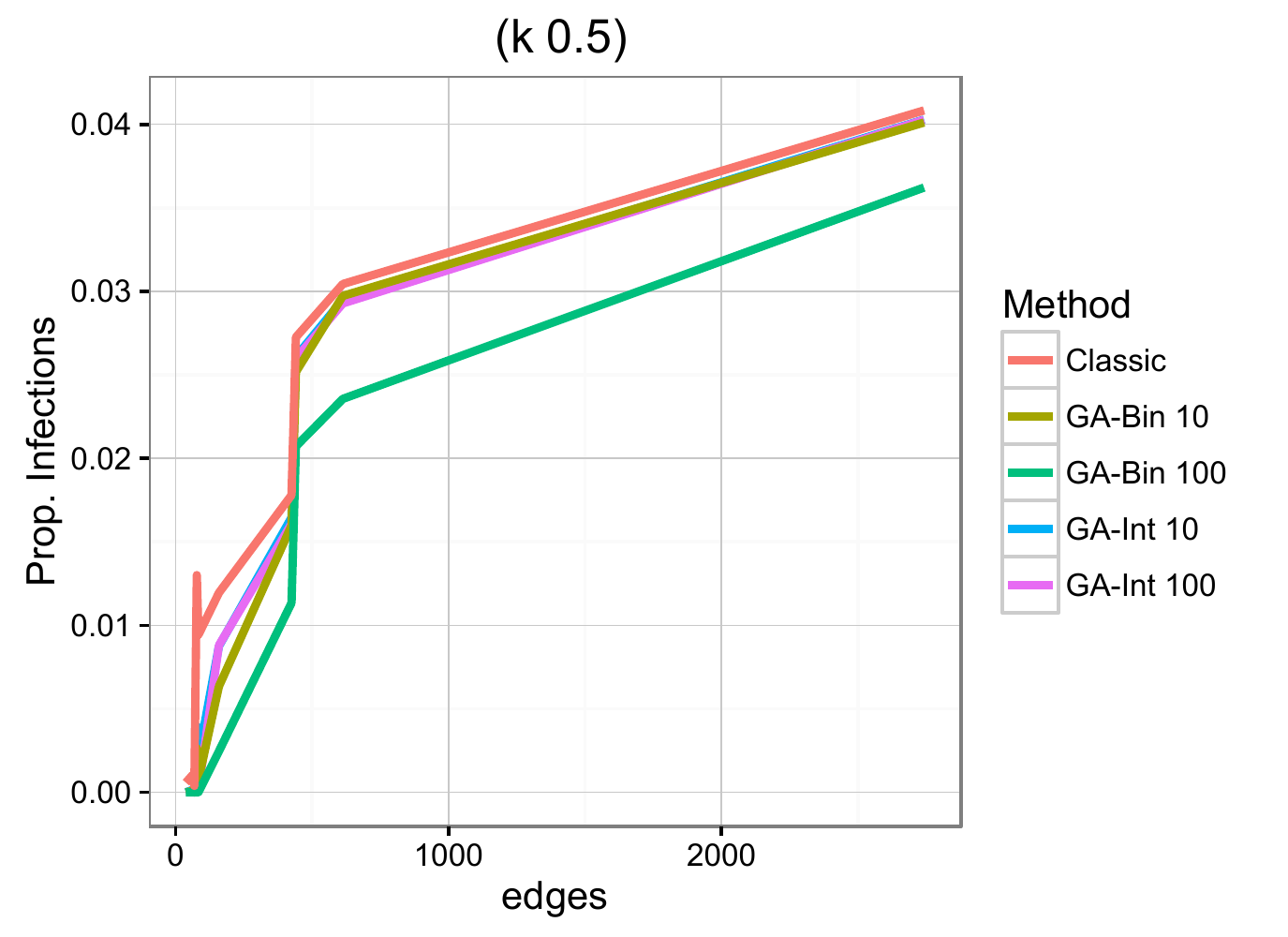}\\
\hline
\end{tabular}
\caption{The scalability of each method and for each $k\in \{0.1\cdot |E|, 0.3\cdot |E|, 0.5\cdot |E|\}$. The proportional number of infections is compared with the number of edges of the tested graphs.}
\label{fig.scalability}
\end{figure*}

In Figure \ref{fig.scalability}, the scalability of the tested methods is shown. The average of the total number of infections is compared with the number of edges from each tested instance. The axis ``Prop. Infections'' is calculated dividing the number of infections by $|V|\cdot 300$, where $|V|$ is the number of nodes from the network, and 300 is the number of iterations/generations. Our genetic algorithms surpassed the Classic heuristic in these results. It can be seen that the Classic heuristic obtained the worst values, and our genetic algorithm GA-Bin 100 obtained the best values. These results suggest that this latter genetic algorithm will find better solutions than the other tested heuristics (especially the Classic method) if the network is larger than the tested ones.

\begin{figure*}[ht]
\centering
\fbox{\includegraphics[width=4.9in]{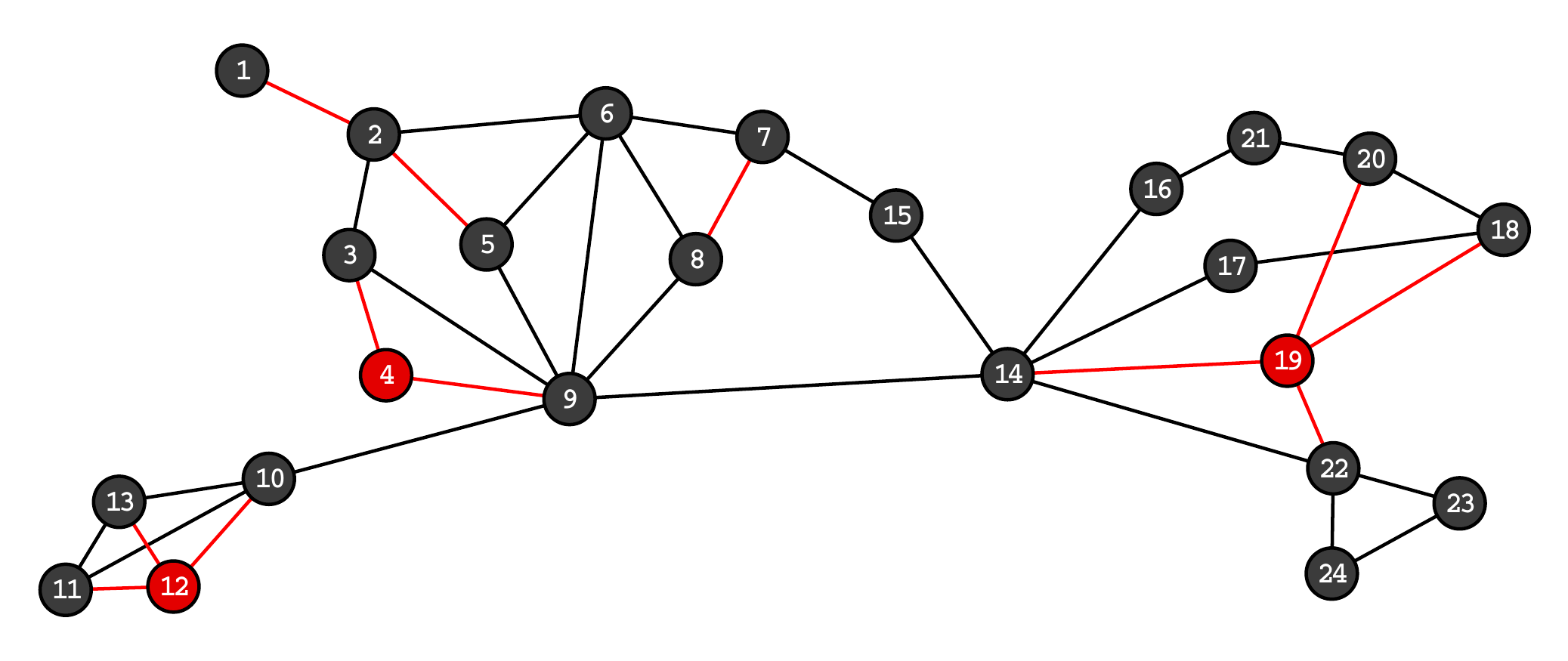}}
\caption{The best solution obtained for Strike when $k=0.3\cdot |E|$.}
\label{fig.sol_strike_03}
\end{figure*}

In Figure \ref{fig.sol_strike_03}, one of the best solutions obtained by our genetic algorithm (GA-Bin 100) when solving the instance Strike with $k=0.3\cdot |E|$ is shown. The red edges are the removed ones (solution), while the red nodes represent the initially infected individuals. In this solution, it is possible to see that the infection spreading was completely contained by isolating the nodes infected in the first period of the simulation time.

\section{Conclusions}

This paper presented a novel genetic algorithm with two different genetic representations that finds solutions for the Min-SEIS-Cluster optimization problem. The problem consists in finding the $k$ edges which minimize the spread of infections in an epidemic, taking into consideration the fact that the chance of transmitting an infection to a node outside of a community is different than if both belong to the same community.

Originally, the Min-SEIS-Cluster optimization problem was solved with a random search. However, the search space for the problem is extremely large, having up to approximately $10^{149}$ possible solutions for an instance with 500 edges. As such, we explored the possibility of utilizing more sophisticated search procedures, with the goal of finding better solutions more quickly. This paper investigated the genetic algorithm metaheuristic, analyzing its effectiveness in solving the Min-SEIS-Cluster problem.

All versions of the genetic algorithm proposed in this paper surpassed the heuristic of \cite{Santiago2016}. The best results are achieved when using the size of the population equal to 100 and binary chromosomes. We called this version ``GA-Bin 100''. In some solutions, the epidemic was completely contained, while in others, the number of infection events along the duration of the epidemic was significantly reduced.

As further investigations, we suggest: (i) to create parallel versions of our genetic algorithms to test larger graphs; (ii) to test other evolutionary strategies and compare them with our results; and (iii) to intensify the search by using the properties of nodes and edges (like centrality) to decide which edges to remove.

\section*{ACKNOWLEDGMENTS}

This work is partly supported by the Brazilian Research Council CNPq, the University of \textit{Vale do Itaja\'i}, and the government of the State of Santa Catarina (Brazil).

\bibliographystyle{ACM-Reference-Format}
\bibliography{percolationlib}


\end{document}